\documentclass[sigconf]{acmart}

\usepackage{setspace}
\usepackage{bbding}
\usepackage{amsmath} 
\usepackage{comment}
\usepackage{todonotes}
\usepackage{pifont}
\usepackage{url}
\usepackage{subfig}

\AtBeginDocument{%
  }

\setcopyright{acmcopyright}
\copyrightyear{2018}
\acmYear{2018}
\acmDOI{XXXXXXX.XXXXXXX}

\acmConference[Conference acronym 'XX]{Make sure to enter the correct
  conference title from your rights confirmation emai}{June 03--05,
  2018}{Woodstock, NY}
\acmPrice{15.00}
\acmISBN{978-1-4503-XXXX-X/18/06}

\begin{document}

\title{DataFlower: Exploiting the Data-flow Paradigm for Serverless Workflow Orchestration}


\author{Zijun Li}
\authornote{Both authors contributed equally to this research.}
\affiliation{%
  \institution{Shanghai Jiao Tong University}
  \city{Shanghai}
  \country{China}
}
\email{lzjzx1122@sjtu.edu.cn}
\orcid{0003-4706-8451}

\author{Chuhao Xu}
\affiliation{%
  \institution{Shanghai Jiao Tong University}
  \city{Shanghai}
  \country{China}
}
\email{barrin@sjtu.edu.cn}
\authornotemark[1]

\author{Quan Chen}
\affiliation{%
  \institution{Shanghai Jiao Tong University}
  \city{Shanghai}
  \country{China}
}
\email{chen-quan@cs.sjtu.edu.cn}

\author{Jieru Zhao}
\affiliation{%
  \institution{Shanghai Jiao Tong University}
  \city{Shanghai}
  \country{China}
}
\email{zhao-jieru@sjtu.edu.cn}

\author{Chen Chen}
\affiliation{%
  \institution{Shanghai Jiao Tong University}
  \city{Shanghai}
  \country{China}
}
\email{chen-chen@sjtu.edu.cn}

\author{Minyi Guo}
\affiliation{%
  \institution{Shanghai Jiao Tong University}
  \city{Shanghai}
  \country{China}
}
\email{guo-my@cs.sjtu.edu.cn}

\renewcommand{\shortauthors}{Zijun Li, Chuhao Xu, et al.}

\begin{abstract}
  Serverless computing that runs functions with auto-scaling is a popular task execution pattern in the cloud-native era.
  By connecting serverless functions into workflows, tenants can achieve complex functionality. 
  Prior researches adopt the control-flow paradigm to orchestrate a serverless workflow. 
  However, the control-flow paradigm inherently results in long response latency, due to the heavy data persistence overhead, sequential resource usage, and late function triggering.
  
  Our investigation shows that the data-flow paradigm has the potential to resolve the above problems, with careful design and optimization. 
  We propose DataFlower, a scheme that achieves the data-flow paradigm for serverless workflows.
  In DataFlower, a container is abstracted to be a function logic unit and a data logic unit. The function logic unit runs the functions, and the data logic unit handles the data transmission asynchronously. Moreover, a host-container collaborative communication mechanism is used to support efficient data transfer.
  Our experimental results show that compared to state-of-the-art serverless designs, DataFlower reduces the 99\%-ile latency of the benchmarks by up to 35.4\%, and improves the peak throughput by up to 3.8X.

  \textbf{Note: This manuscript is our major  version, and is accepted after revision in ASPLOS 2024. }
\end{abstract}

\received[accepted after revision]{28 April 2023}

\maketitle

\section{Introduction}

Serverless computing is popular in the cloud-native era~\cite{DBLP:journals/usenix-login/HendricksonSOHV16,DBLP:journals/corr/abs-1902-03383,DBLP:journals/corr/abs-2112-12921,DBLP:conf/micro/ShahradBW19,DBLP:conf/usenix/WangLZRS18}, and cloud computing vendors all provide serverless computing services (e.g., AWS Lambda~\cite{lambda}, Microsoft Azure Functions~\cite{azurefunctions}, and Alibaba Function Compute~\cite{alifc}). 
When a user submits a request with serverless computing, the corresponding function is triggered to run in a container. 
As the logic of a function is simple, a complex application is usually implemented as a function workflow. 
Traditionally, programmers build such a workflow by writing the code to trigger each function manually, and they also need to handle function orchestration, intermediate data transfer, and error retry~\cite{lambdadrawbacks,Netherite,DBLP:journals/pacmpl/JangdaPBG19}.
Cloud computing vendors therefore offer an easier way to build and run a function workflow, called serverless workflow. 
With serverless workflow, programmers only need to declare the call dependencies between functions. The serverless platform automatically manages function orchestration, intermediate data, and fault tolerance~\cite{DBLP:conf/asplos/JiaW21,DBLP:conf/usenix/KlimovicWKSPT18,DBLP:conf/usenix/KotniNGB21,DBLP:conf/asplos/LiLGCCZG22,DBLP:conf/osdi/MahgoubYSECB22,DBLP:journals/pomacs/MahgoubYSMEBC22}.

\begin{figure}
  \centerline{\includegraphics[width=.85\columnwidth]{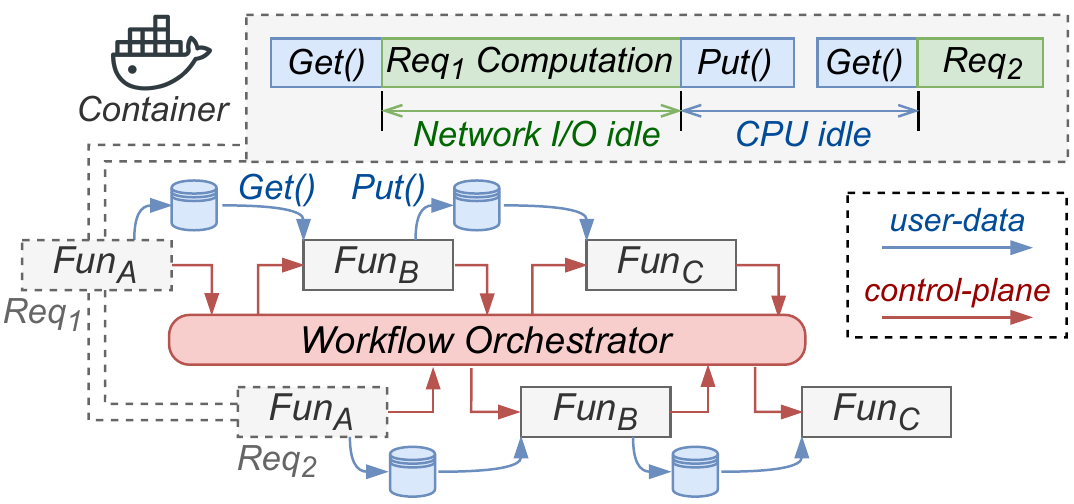}}
  \caption{\label{fig:control-flow} Run a workflow using the control-flow paradigm. An orchestrator decides when/where to trigger a function.}
  \vspace{-2mm}
\end{figure}

For a serverless workflow, based on its call dependency graph, serverless platforms (e.g., AWS Step Functions~\cite{StepFunctions}, Alibaba Serverless Workflow~\cite{aliworkflows}, OpenWhisk~\cite{OpenWhisk} and Fission Workflows~\cite{Fissionwf}) use a workflow orchestrator (or controller) to trigger a function based on the completion status of its predecessor functions. 
As shown in Figure~\ref{fig:control-flow}, the orchestration is typically done based on the control-flow paradigm. 
In general, the orchestrator maintains the states of all functions and triggers functions in a sequential order based on user-defined control dependencies. 
When all predecessors of a function are complete, the orchestrator triggers it for execution.
Once triggered, the function's input data is loaded from the backend storage. If a function needs to transfer output data to a destination function, the sender stores the intermediate data in backend storage (mainly remote storage) and the receiver reloads it~\cite{DBLP:conf/usenix/BoucherKAK18,DBLP:conf/usenix/FouladiRILCKZW19,DBLP:conf/cidr/HellersteinFGSS19,DBLP:conf/sigmod/0002MA20}. 
The persistence of intermediate data is necessary for the mainstream stateless serverless computing. 
Some platforms~\cite{StepFunctions,durable} also provide workflow interfaces for stateful functions. They support the persistence of small data without backend storage (i.e. $<$256KB in AWS Step Functions~\cite{stepquota}). Larger data still relies substantially on backend storage.

Analyzing the above steps, the orchestration based on the control flow paradigm suffers from three main limitations. 
Firstly, the double transfer of the intermediate data from the source to the backend storage, and from the storage to the destination results in heavy communication overhead. Especially in the case that a serverless container has limited network bandwidth for fairness in production~\cite{10.1145/3490386,DBLP:conf/usenix/WangLZRS18}.
Secondly, by default, the orchestrator does not allow a container to run multiple function invocations concurrently~\cite{lambdaconcurrency,DBLP:conf/nsdi/AgacheBILNPP20,DBLP:conf/cloud/DukicBSA20,DBLP:conf/asplos/JiaW21}.
The sequential steps of executing a function (i.e. loading data with Get(), executing the function, storing data to backend storage with Put()) will idle either the CPU or the network (the upper part of Figure~\ref{fig:control-flow}). Resources are not fully utilized. 
Thirdly, the state management overhead and the state-driven triggering pattern of the orchestrator make the actual trigger time of a function later than its ready time.

The data-flow paradigm~\cite{DBLP:conf/isca/DennisM74,DBLP:journals/corr/abs-2112-12921,DBLP:journals/fgcs/Ramon-CortesLEB20} 
has the potential to properly resolve the above issues.
Following the data-flow paradigm, 
a data-flow graph describes the data dependencies between functions, and a function starts to run immediately once all its input data is ready.
Based on the data-flow graph, the intermediate data flows asynchronously to the destination as soon as the data is generated, without relying on backend storage. This feature allows computation and communication to overlap. 
Moreover, each function determines whether it can start to run independently without a centralized orchestrator.

It is nontrivial to use the data-flow paradigm in serverless workflows.
Most importantly, the computation and data transmission logic are all tightly coupled in the user function code. The current serverless frameworks do not abstract data-flow graphs for workflows. 
It is necessary to redesign the serverless programming and execution model to separate a function's computation logic and the data transmission logic. In this way, mechanisms can be designed and implemented to overlap the compu./comm. 

Secondly, the workflow orchestrator has to be re-architected to support the data-flow based function triggering.
As serverless computing is typically applied to stateless applications, some function containers are generated/recycled, and the data-flow graph changes dynamically.
A decentralized workflow engine, capable of efficiently synchronizing the up-to-date data-flow graph, is essential to ensure the correct communication and execution results.


Thirdly, the direct communication mechanism between functions without persistent storage should be carefully designed. 
The destination container may not even be started when the data flows to it from the source function, due to the cold startup and time-consuming user-related environment setup.
A data cache and the corresponding cache management mechanism are necessary on each host node to support the low-overhead data transfer.

Some prior frameworks also use data-flow paradigm to schedule task workflows~\cite{Flink,tensorflow,DBLP:journals/fgcs/DeelmanVJRCMMCS15,DBLP:conf/hotcloud/ZahariaCFSS10}.
They are not applicable to serverless workflows because they orchestrate the tasks assuming that all the tasks in the workflow are addressable beforehand. However, a function container is not addressable beforehand in serverless.
Moreover, they assume that at least one executor container is alive and ready to communicate. This assumption is not the case with serverless as well.
Lastly, prior researches do not consider the limited memory space and network bandwidth of a container in serverless.

We therefore propose {\bf DataFlower}, a data-flow paradigm-based serverless workflow scheme. 
In terms of the execution model, we abstract the container for a function as {\it Function Logic Unit (FLU)} and {\it Data Logic Unit (DLU)}. The FLU is responsible for executing the function code, and the DLU transforms the output data from the FLUs and sends it to the destination.
On each node, a {\it workflow scheduling engine} parses the data-flow graph, and synchronizes the data-plane with DLUs. The engine replaces the workflow orchestrator to scale containers for functions, based on the transmission pressure on each DLU.
In addition, DataFlower introduces a {\it host-container collaborative communication mechanism} to support the implicit data transfer between functions. A streaming-based pipe connector is used to minimize the data transfer latency.
On each host node, a data sink is maintained for each function to temporarily hold its data.
The data in the sink automatically expires once the data is loaded by the container, to reduce the memory overhead.



We have implemented a serverless framework based on the DataFlower scheme, and the framework is open-sourced with a hidden name. To the best of our knowledge, 
DataFlower is the first work that uses the data-flow paradigm to orchestrate serverless workflows.
The main contributions are as follows.


\begin{itemize}
\item \textbf{The semantic comparison and analysis of control-flow and data-flow paradigms for serverless workflows.} The investigation shows the drawbacks of using the control-flow paradigm for scheduling serverless workflows, and highlights the advantages of the data-flow paradigm.
\item \textbf{The scheme of a data-flow driven function orchestration for serverless workflow.} It enables autonomous function-to-function communication, early function triggering, and computation-communication overlap in a container.


\item \textbf{A set of mechanisms to improve the data-flow driven orchestration for serverless workflows.} We reveal that the communication backpressure and inefficient data lifetime management are performance bottlenecks. Pressure-aware function scaling and host-container collaborative mechanisms are proposed to tackle the bottlenecks.


\end{itemize}

We evaluate DataFlower using real-world applications. Extensive experimental results show that compared to control-flow based serverless designs, 
DataFlower reduces the 99\%-ile latency of applications by up to 35.4\% and improves the peak throughput by up to 3.8X, while reducing resource usage by up to 69.3\%.


\section{Related Works}
\label{sec:related}

\textbf{Data-passing in serverless workflows.}
GlobalFlow~\cite{DBLP:conf/IEEEcloud/ZhengP19} proposed a connector-based strategy to effectively coordinate logically dependent serverless services on geographically distributed scenarios. SONIC~\cite{DBLP:conf/usenix/MahgoubSMKCB21} improved the transmission efficiency of different fan-out branches by introducing an application-aware data-passing mechanism between functions in the workflow. 
Diverse data-passing and storing solutions for serverless functions~\cite{DBLP:conf/sosp/JiaW21,DBLP:conf/osdi/KlimovicWSTPK18,DBLP:conf/nsdi/PuVS19, DBLP:journals/vldb/WuSH21} further optimized the data I/O bottlenecks.
Though the data transfer overhead can be partially reduced, the late function triggering due to the control-flow paradigm still exists. 

\textbf{Task-based serverless workflows.}
Most serverless designs adopt the centralized workflow orchestrator to maintain the execution state, and then use the control-flow to assign tasks~\cite{DBLP:journals/fgcs/Balis16,DBLP:conf/asplos/FuerstS21,DBLP:journals/fgcs/MalawskiGZBF20,DBLP:conf/asplos/JiaW21,DBLP:conf/cloud/TariqPNRL20}. To reduce the scheduling overhead, NightCore~\cite{DBLP:conf/asplos/JiaW21} used lightweight IPC for functions co-located on the same node. For cross-node function triggering, it relies on the gateway to collect the execution state and assign function tasks. FaaSFlow~\cite{DBLP:conf/asplos/LiLGCCZG22} proposed a decentralized scheduling pattern called WorkerSP to reduce the cross-node scheduling overhead. Other related works like function bundling of a workflow~\cite{DBLP:conf/usenix/AkkusCRSSBAH18,DBLP:conf/usenix/LiG0CXZSMY0G22,DBLP:journals/pomacs/MahgoubYSMEBC22,DBLP:conf/osdi/MahgoubYSECB22, DBLP:conf/usenix/KotniNGB21,DBLP:conf/asplos/JiaW21,DBLP:conf/usenix/ShillakerP20} explored the benefits when hosting multiple functions of an application into the dedicated sandbox. They still focused on the traditional control-flow based serverless frameworks, while DataFlower is a data-flow paradigm-based serverless scheme whose function triggering is based on the data-availability. 

\textbf{Dataflow-based workflows.}
Pegasus~\cite{DBLP:journals/fgcs/DeelmanVJRCMMCS15} is a DAG (Directed Acyclic Graph) based workflow system, its trigger logic depends on the completion state of each task. Spark~\cite{DBLP:conf/hotcloud/ZahariaCFSS10} focuses on big data processing and offers data-oriented optimizations such as data-locality aware scheduling. Other similar systems such as Flink~\cite{Flink} and TensorFlow~\cite{tensorflow} are also dataflow-based workflow systems. However, they hold the assumptions that 1) tasks can communicate with each other through addressable connections, 2) executor containers typically have large local memory, and 3) at least one executor container is alive and ready to communicate. These assumptions do not apply to serverless.

To enable the data-flow paradigm, SWEEP~\cite{DBLP:conf/ucc/JohnAMKT19} implemented a workflow management layer on top of serverless frameworks to enable the definition and scalable execution of data processing pipelines.
AFCL~\cite{DBLP:journals/fgcs/RistovPF21} proposed a high-level programming abstraction to express control-flow and data-flow when constructing a serverless workflow, based on the AWS Lambda backend. However, without co-optimization with the underlying serverless architecture and a customized container scheduling strategy, they cannot take full advantage of the data-flow paradigm like DataFlower in serverless.

\section{Serverless Workflow Paradigm Investigation}
\label{sec:moti}
In this section, we first explain the background of serverless workflow.
We then discuss the control flow paradigm for serverless computing. Finally, we will analyze the potential of using the data-flow paradigm to alleviate the drawbacks.

\begin{figure*}
  \captionsetup[subfloat]{margin=2pt,format=hang}
  \centering
  \subfloat[End-to-end communication and computation time breakdown]{\includegraphics[width=.444\textwidth]{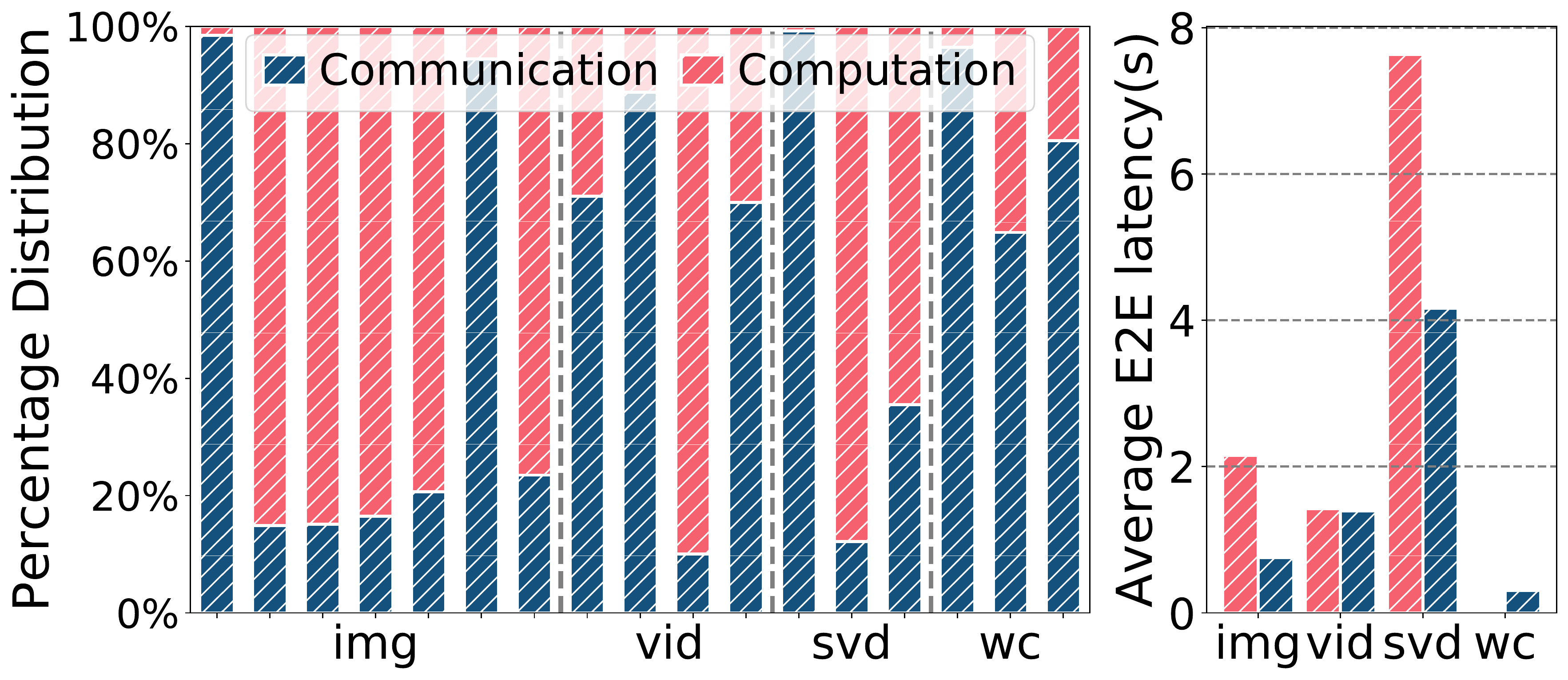}}
  \subfloat[Resource usage timeline]
  {\includegraphics[width=.32\textwidth]{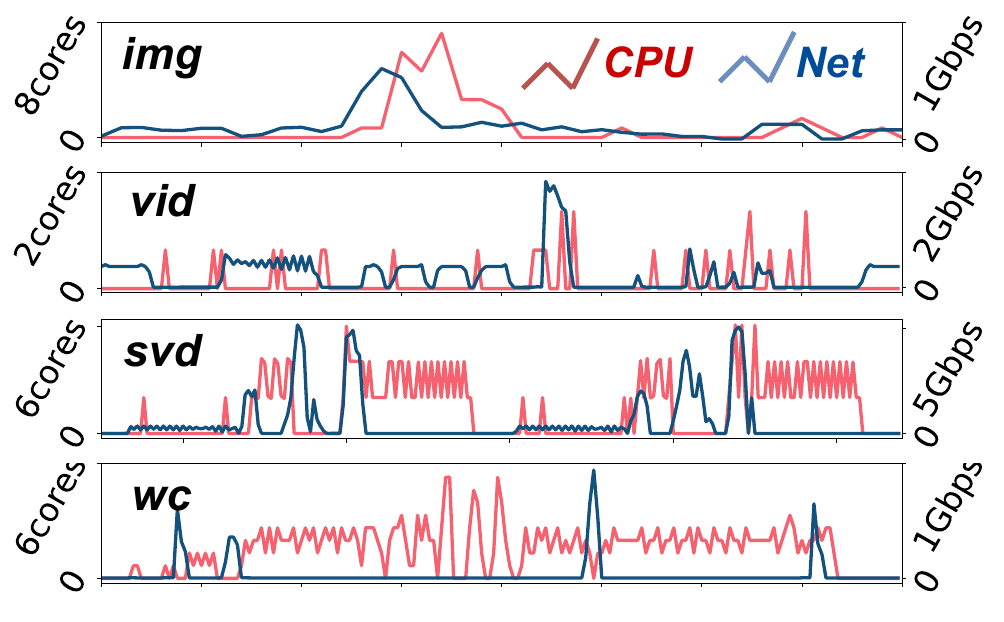}}
  \subfloat[Triggering overhead]
  {\includegraphics[width=.227\textwidth]{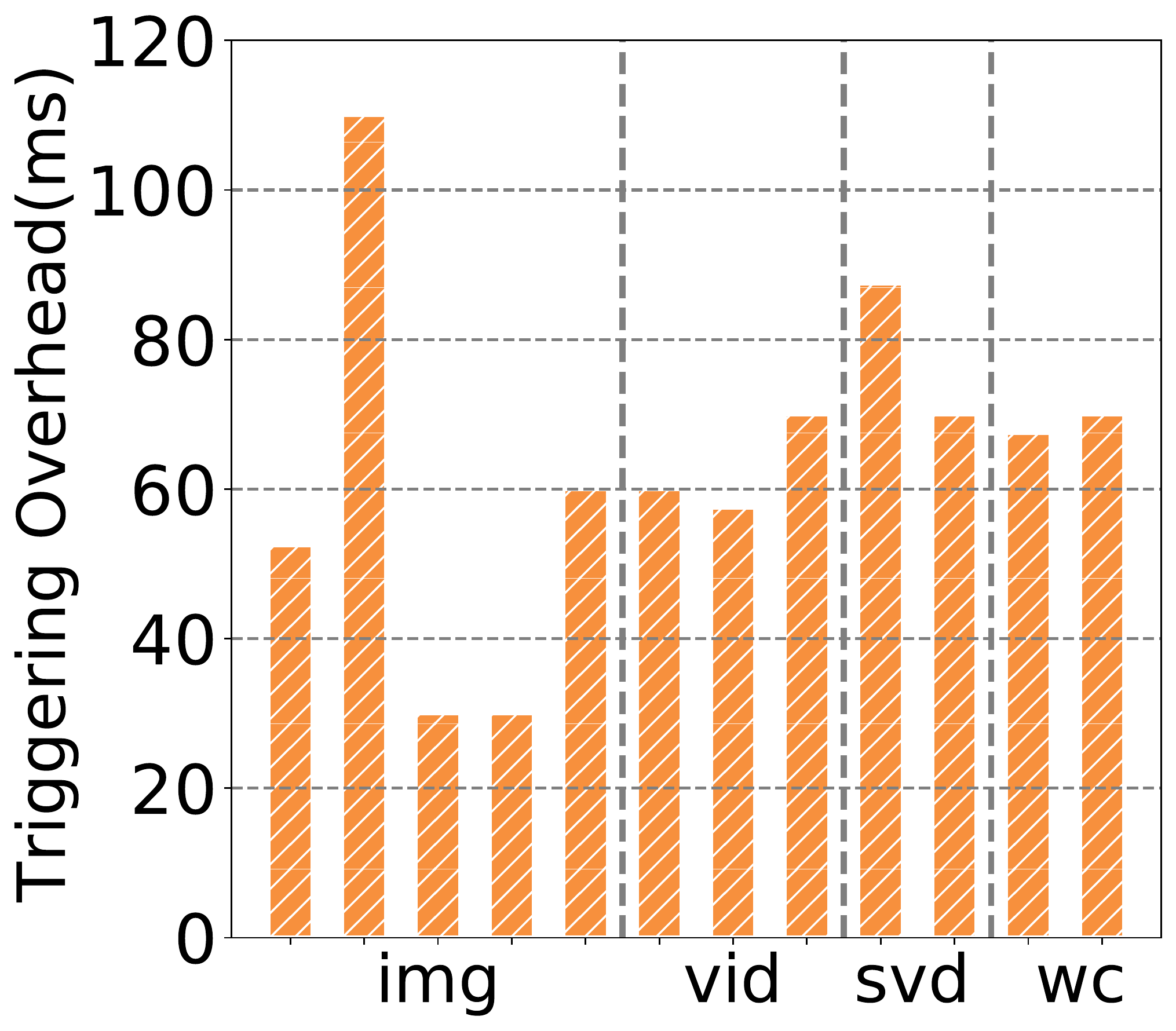}}
  \caption{\label{fig:iologic} The e2e latency breakdown, resource usage timeline, and state management overhead with the control-flow paradigm.}
\end{figure*}

\subsection{Backgrounds of Serverless Workflows}
In the cloud-native era, a complex application is often decoupled into fine-grained functions. 
Each function scales independently with serverless computing. 
Many popular applications~\cite{DBLP:conf/cloud/AoIVP18,DBLP:conf/nsdi/FouladiWSBZBSPW17,DBLP:conf/cloud/CarreiraFTZK19} have already been decoupled and orchestrated to be a serverless workflow, to achieve the functionality of the original application~\cite{DBLP:conf/asplos/JiaW21,DBLP:journals/csur/AdhikariAS19,DBLP:conf/asplos/LiLGCCZG22,DBLP:conf/osdi/MahgoubYSECB22,DBLP:conf/usenix/KotniNGB21,DBLP:journals/pomacs/MahgoubYSMEBC22}. 

The existing workflow definition relies substantially on the Workflow Definition Language to describe the control and dependency logic between functions within a workflow, while modeling it as a state machine to templatize the trigger logic on the control-plane~\cite{cncf,awsstate}. In general, a serverless workflow can be represented by a DAG (Directed Acyclic Graph)~\cite{DBLP:journals/csur/AdhikariAS19,DBLP:conf/cloud/BhasiGTMKD21,DBLP:conf/cloud/CarverZWAWC20,DBLP:journals/tpds/LinK21,DBLP:conf/cloud/TariqPNRL20,DBLP:conf/IEEEcloud/ZhengP19}, where the nodes represent the functions and the edges represent the control dependencies.

When a serverless workflow request arrives, workflow functions are triggered according to dependencies.
If all predecessors of a function are complete, the orchestrator triggers it to run.
If there is already a warm container for the function, it reuses the warm container directly. 
Otherwise, a new container is started to host the function, incurring the overhead of a cold startup~\cite{DBLP:conf/icdcsw/McGrathB17, DBLP:conf/usenix/WangLZRS18, DBLP:conf/nsdi/PuVS19, DBLP:conf/micro/ShahradBW19, DBLP:conf/usenix/LiG0CXZSMY0G22,DBLP:conf/asplos/DuYXZYQWC20,DBLP:conf/eurosys/WangHW19,DBLP:conf/asplos/Ustiugov0KBG21}.

\subsection{The Control-flow Paradigm for Serverless Workflows}

Current serverless platforms~\cite{StepFunctions,Fissionwf,OpenWhisk,DBLP:conf/usenix/KotniNGB21,DBLP:conf/asplos/LiLGCCZG22,DBLP:conf/usenix/MahgoubSMKCB21,DBLP:conf/asplos/JiaW21} adopt the control-flow paradigm to model and execute a serverless workflow.
With control-flow, the data plane ensures the correctness of communication, and the control plane controls the invocation time and order of the functions.

With the control-flow paradigm, the generated data by a completed function should be persisted in a backend storage or passed to the destination function through the message queue.
It is necessary to persist the intermediate data, as users are not aware of the data locality and containers may be recycled once timeout~\cite{DBLP:conf/usenix/MahgoubSMKCB21,DBLP:conf/usenix/BoucherKAK18,DBLP:conf/nsdi/PuVS19,DBLP:conf/sigmod/0002MA20,DBLP:conf/usenix/FouladiRILCKZW19}.
On the control-plane, 
a workflow orchestrator monitors the states of the functions,
and decides when and where to trigger the next function 
based on the control dependencies and the data-availability.

We first run serverless workflows on 3 popular production serverless platforms (i.e.,  AWS Step Functions, Azure Durable Functions, Alibaba Serverless Workflows) to study their performance. 
These platforms use the control-flow paradigm to orchestrate a serverless workflow. 
Four production-level best practice serverless workflows, \textit{Video-FFmpeg} ({\it vid})~\cite{ffmpeg}, \textit{ML-based Image Processing} ({\it img})~\cite{imagemagick}, \textit{Singular Value Decomposition} ({\it svd})~\cite{DBLP:journals/pomacs/MahgoubYSMEBC22} and \textit{WordCount} ({\it wc})~\cite{DBLP:conf/icde/ZhangHDZH17}, are used as the benchmarks to perform the investigation. 
The benchmarks cover the same serverless scenarios as in FaaSFlow~\cite{DBLP:conf/asplos/LiLGCCZG22} and WISEFUSE\footnote{The benchmarks in WISEFUSE are not open-sourced currently.}~\cite{DBLP:journals/pomacs/MahgoubYSMEBC22}.
As an example, Figure~\ref{fig:iologic} shows the characterization result on one of the serverless platforms. Other production platforms show similar characteristics.
 
\subsubsection{Heavy Data Persistence Overhead.}
Figure~\ref{fig:iologic}(a) shows the computation and communication time breakdown of the benchmarks in the production serverless platform. In the figure, each bar represents a function in the workflow. The communication time consists of the time to retrieve the data from the backend storage and the time to store the data in the backend storage. 
We can observe that the communication time is dominant for many functions in all the four serverless workflow benchmarks~\cite{DBLP:journals/corr/abs-2109-13492}. 
For the four benchmarks, communication accounts for 26.0\%, 49.5\%, 35.3\% and 89.2\% of the end-to-end latency, respectively. One recent research also reveals the communication bottleneck in production environments~\cite{DBLP:conf/usenix/KlimovicWKSPT18}. 
The communication overhead would be even larger, if more concurrent or bursty queries from the same function use the backend storage concurrently.


The communication time is long due to two main reasons. First, the data is transferred twice (from the source function to the backend storage, and from the storage to the destination function). 
Moreover, backend storage still has high access latency and offers
either limited I/O performance, especially for highly scalable serverless containers from a single function.
The double data transfer and the limited network bandwidth result in the heavy data persistence overhead.

These platforms that allow stateful functions to communicate without remote storage can eliminate the data persistence overhead for the small intermediate data. 
For instance, the state machine in AWS Step Functions supports the data cache smaller than 256KB~\cite{stepquota}. Large intermediate data is still persisted through the remote backend storage. They also suffer from the data persistence overhead when executing our benchmarks and other large-scale serverless workflows.

\subsubsection{Sequential Resource Usage.}


Figure~\ref{fig:iologic}(b) shows the CPU and network usage timeline when running these benchmarks. 
In the experiment, we collect the resource usage on each container. The figure reveals 
the CPU and network usage of all running functions in a benchmark. 
The functions run sequentially in this experiment.
As observed, either the CPU core or the network of a container is staggered to peak usage during its lifetime.
The network I/O resource is mainly consumed during the data access phase (Get() and Put()), while the CPU is waiting for the I/O completion, and the thread is suspended. Similarly, when running the computation logic, the network is idle.
Sequential resource usage is inherent, 
as the computation and data transmission logic are tightly-coupled in the user code with the control-flow paradigm. 
If multiple invocations of a function happen, the invocations queue up.



\subsubsection{Late Function Triggering.} 

The functions are triggered sequentially because the workflow orchestrator
automatically enables the sequential execution of diverse logics (i.e, parallel, switch, and foreach). 
To maintain the correct order when processing these functions, the workflow orchestrator schedules and triggers the functions in a workflow sequentially in the topological order.
This in-order triggering pattern prevents a function from being triggered early, even if all its input data is ready.
Experiences in the computer architecture field have shown that out-of-order execution improves performance.


Figure~\ref{fig:iologic}(c) shows the time needed to manage the function states for ordering the function triggering on the control-plane with the production orchestrator. The triggering overhead is calculated to be the duration between the end timestamp of a function and the start timestamp of its successive function in the log file.
As observed, it takes 63.3 milliseconds on average to manage the states between two adjacent functions.
The overhead is relatively long, when a function's execution time is often short for serverless computing.

The three drawbacks of the control-flow paradigm originate from its inherent logic, and thus cannot be alleviated through minor modification. For instance, FaaSFlow~\cite{DBLP:conf/asplos/LiLGCCZG22} uses local memory to cache data for local functions, and SONIC~\cite{DBLP:conf/usenix/MahgoubSMKCB21} caches the data in the source function and the destination function fetches the data when it is triggered. However, they all suffer from sequential resource usage and late function triggering with the control-flow paradigm. 
The data is not transferred to the destination beforehand. 
We also show the performance of FaaSFlow and SONIC in Section~\ref{sec:eval}.




\subsection{The Potential Advantages of Data-flow Paradigm}

The data-flow paradigm that defines the data dependencies explicitly 
is capable of enabling out-of-order execution and maximizing the parallelism.
The data dependencies are expressed to be a data-flow graph in the data-flow paradigm.
For each function, the data-flow graph records the source functions of its inputs, and the destination functions of its outputs.

\begin{figure}
  \centerline{\includegraphics[width=\columnwidth]{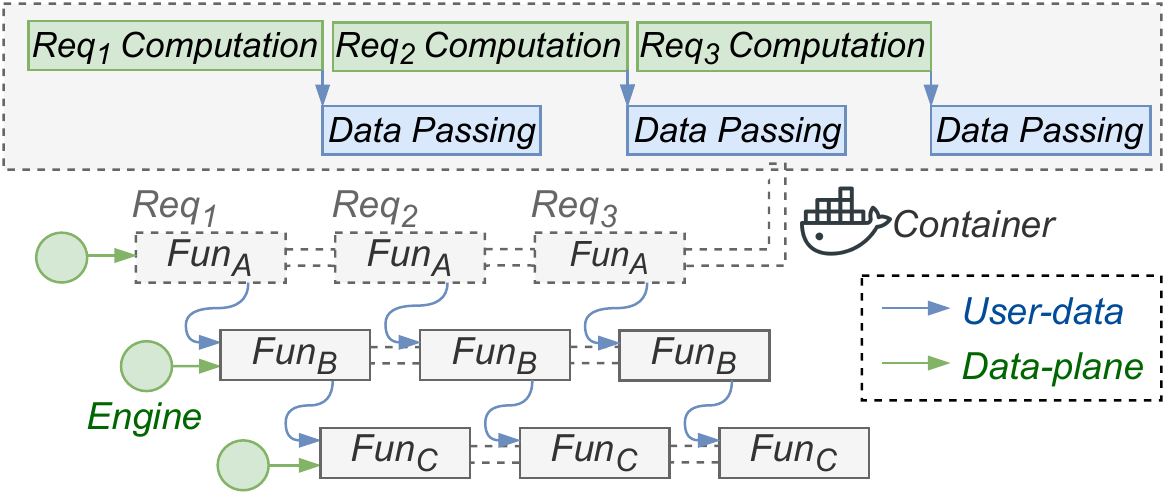}}
  \caption{\label{fig:sec2-3} Run a workflow using the data-flow paradigm. Data flows to destinations according to the data-flow graph.}
\end{figure}

Figure~\ref{fig:sec2-3} illustrates the better way we proposed to run a serverless workflow using the data-flow paradigm~\cite{DBLP:conf/isca/DennisM74,DBLP:journals/fgcs/Ramon-CortesLEB20,DBLP:journals/corr/abs-2112-12921}. 
The basic idea here is decoupling and processing the logic of computation and communication independently.
In this way, the data-passing operations may be done asynchronously when the container executes the function,
and a container can run the next function invocation before the data passing completes.
The data-flow paradigm resolves the drawbacks as follows.

\subsubsection{Alleviating Data Persistence Overhead.}
With the up-to-date data-flow graph, a function knows destinations of its outputs. 
It is possible to directly transfer the data to destinations without backend storage.
Moreover, with direct transfer, the streaming technique can be introduced. 
When a data chunk is ready, it can be transferred to the destination before the entire data is generated.
The streaming technique can be implemented with a pipeline connector that connects the source and the destination.

\subsubsection{Improving Resource Utilization.}



As shown in the upper part of Figure~\ref{fig:sec2-3}, a container 
can execute a new request before the data transfer of the previous one completes.
In this way, the computation and the communication overlap, and resource utilization is improved.
The overlap is beneficial for reducing the response latency of a workflow at a high load,
as the queries may start earlier.


\subsubsection{Early Function Triggering.}


The data-flow paradigm enables the early function triggering from two aspects, as shown in the lower part of Figure~\ref{fig:sec2-3}.
As for the first aspect, it enables out-of-order function triggering. 
As for the second aspect, the data of a function can start to be transferred to the destination functions earlier with the data-flow paradigm, and the destination's input data is ready earlier.
It is possible that the data is generated in the middle of the function, but is transferred to the destination after the function completes, using the control-flow paradigm. 

Based on the above analysis, we have two main insights. Insight 1: The widely-used control-flow paradigm inherently fails to minimize the end-to-end latencies and maximize the throughput of serverless workflows. 
Insight 2: Data-flow paradigm 
enables the design of reducing the data transfer overhead, overlapping compute and data transfer, 
and early function triggering and execution.

\subsection{Challenges of Applying Data-flow Paradigm}
Designing a serverless workflow scheduling scheme based on the data-flow paradigm faces three main challenges.

{\bf Challenge-1: The computation and data passing should be decoupled}, however current serverless frameworks do not support such operation. 
The programming interface and the execution model should be re-designed to support the explicit 
data-flow graph declaration, and enable the ability to perform the computation and data transmission asynchronously. 


{\bf Challenge-2: The workflow engine should fundamentally support the decentralized function triggering based on data-availability}. 
Specifically, the workflow engine is required to parse the data-flow graph, monitors the availability of functions' input data, and dynamically schedules available containers for function invocations accordingly.


{\bf Challenge-3: The direct data communication mechanism should be carefully designed to support implicit data-plane.} When a function transfers its data to the destination, the destination function has not yet been triggered and no container can receive the data. Therefore, the host node of the destination function should properly manage the data before the function is triggered, and recycle the data to reduce the memory overhead.

\section{The Design Methodology of DataFlower}
\label{sec:dataflower}

\begin{figure}
  \centerline{\includegraphics[width=\columnwidth]{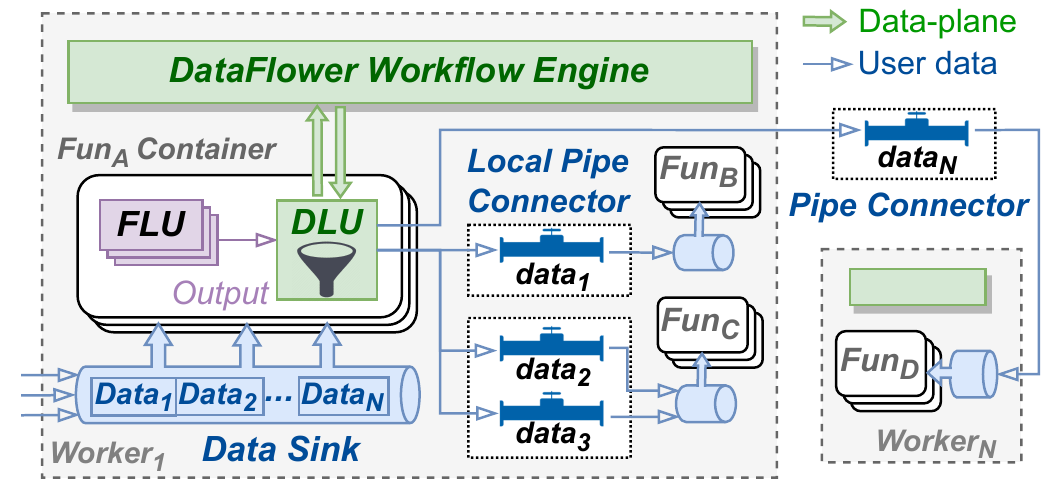}}
  \caption{\label{fig:sec3} The scheme overview of DataFlower.}
\end{figure}
 
Figure~\ref{fig:sec3} presents {\bf DataFlower}, a scheme that achieves the data-flow paradigm for the serverless workflows.
As shown, in each physical node, DataFlower deploys a {\it data-flow based scheduling engine} to schedule all the functions on the node. 
A container for a function is abstracted into a {\it function logic unit} (FLU), and a {\it data logic unit} (DLU).
The FLU runs the function, and the DLU receives the input data for the function, and pushes the output data to the destination functions.
Moreover, between each source and destination function pair, a {\it pipe connector} is used to transfer the intermediate data.

DataFlower runs a serverless workflow in following steps. 
First, once the workflow is invoked, DataFlower relies on the default function mapping method to assign functions to different nodes. DataFlower also provides an open interface for the function mapping method, so that other function mapping solutions can be implemented.
Second, the data-flow graph is informed to the workflow engine on each node.
Third, according to the data-flow graph, the workflow engine monitors whether each function's input data is ready. If all the input data of a function is ready, 
the engine triggers the function, and finds it an appropriate container to run it. 
Fourth, the container uses the FLU to run the function, and the DLU of the container writes the output of the function to the destination nodes, according to the data-flow graph, using the pipe connector.

Note that, when the data is ready to be transferred to the destination function, it is possible that the container of the destination function has not been set up yet. In this case, the transmission is blocked.
To this end, as shown in Figure~\ref{fig:sec3}, DataFlower designs a {\it host-container collaborative communication mechanism} that caches the data in the {\it data sink} on every host node. The data sink of a node accepts and caches the data from all the other nodes, so that a function is able to quickly obtain its input data, once it is triggered to run. 


Due to the time-consuming data transfer per container, it is also possible that the DLU pushes the data through the pipe connector at a much slower rate than the data generated into the DLU.
In this case, it is better to scale out more containers for the function to avoid the backpressure problem.
We further propose a {\it pressure-aware function scaling mechanism} which monitors the data stream pressure on the DLU and determines whether to scale out the function containers.
\section{The FLU and DLU Abstraction}
\label{sec:fluabstraction}
In this section, we first introduce the programming model that supports the FLU and DLU abstraction. 
Then, we present a pressure-aware function scaling mechanism that enables the container auto-scaling for each function in a workflow.

\subsection{FLU-DLU Programming and Execution Model}

With the data-flow paradigm, the function computation logic and the data transmission logic should be separated to enable the potential computation and transmission overlap.
However, the current workflow programming languages for serverless computing do not support the separation. 

\begin{figure}
  \captionsetup[subfloat]{margin=5pt,format=hang}
  \vspace{-3mm}
  \centering
  \subfloat[Pseudocode of separating FLU and DLU.]{\includegraphics[width=.21\textwidth]{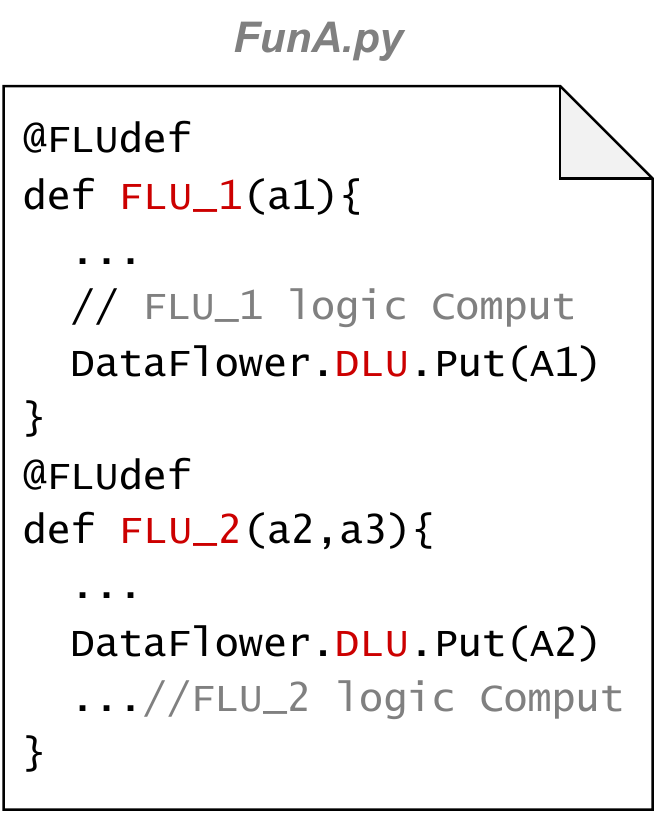}}
  \subfloat[Example of the FLU triggering and DLU interaction.]{\includegraphics[width=.27\textwidth]{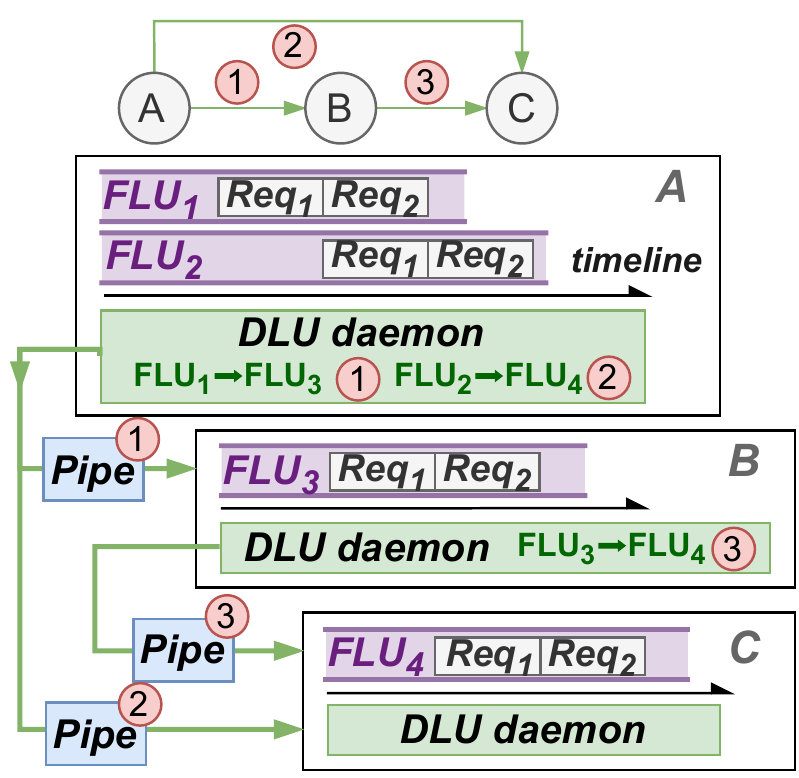}}
  \caption{\label{fig:interaction} The programming and execution model that supports the FLU and DLU abstraction.}
\end{figure}

Figure~\ref{fig:interaction}(a) shows the programming model in DataFlower that supports the computation and data transmission separation. As shown, DataFlower provides an interface for the function to declare the data to be handled by the DLU. The FLU then executes the function and the DLU receives data from the function FLU and transfer data to the destinations. 
A FLU starts when a function is invoked in the container, and the DLU runs as a daemon process in the background.

It is possible for a function to generate data for multiple child functions (e.g. parallel, foreach). 
For simplicity, one possible design is to use the DLU to handle all the data at the end of the function.
This design results in the late transfer, as some data may already be ready in the middle of the function. 
To this end, in the DataFlower programming model, whenever the data for a function is ready, the DLU should be called to initiate the transmission immediately. If a FLU generates data that has multiple optional destinations (i.e. switch), its DLU is able to select the correct destination based on the user-defined data-flow logic. Therefore, DataFlower naturally supports dynamic DAG declarations~\cite{DBLP:conf/cloud/BhasiGTMKD21}.

The DLU should be called at least once in each FLU definition to ensure that the intermediate data can flow out and trigger the next function FLU. 
For the terminal FLU function, an $end$ signal is required to exit even if a user does not need a returned result. In addition, the computation of a function can be divided into multiple FLUs to increase the intra-function execution parallelism. Such abstraction further empowers the pipeline execution of the FLUs within a container.


Figure~\ref{fig:interaction}(b) shows the execution of a data-flow graph with DataFlower.
As shown, the FLUs and the DLU of a container run asynchronously, and multiple FLUs in a function can run in a pipeline to maximize the parallelism. 
The DLU of a container on a node collaborates with the node's engine to transfer data. 
For each workflow request, a data-plane is provided by the scheduler and loaded into the DLU daemon to locate the destination functions (e.g. IP address, socket ID). 
$FLU_3$ can be triggered early when data arrives from $FLU_1$, without the necessity of waiting for $Fun_A$ to complete.

\subsection{Pressure-aware Function Scaling}
\label{sec:pressure_scaling}
 \begin{figure}
    \centering
    \subfloat[Backpressure occurs if FLU produces more than DLU consumes.]
    {\includegraphics[width=.44\textwidth]{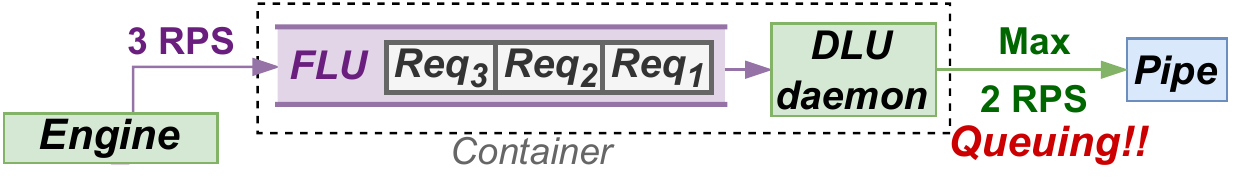}}
    \vspace{-2mm}
    \subfloat[Pressure-aware notification for FLU-DLU scheduling.]{\includegraphics[width=.44\textwidth]{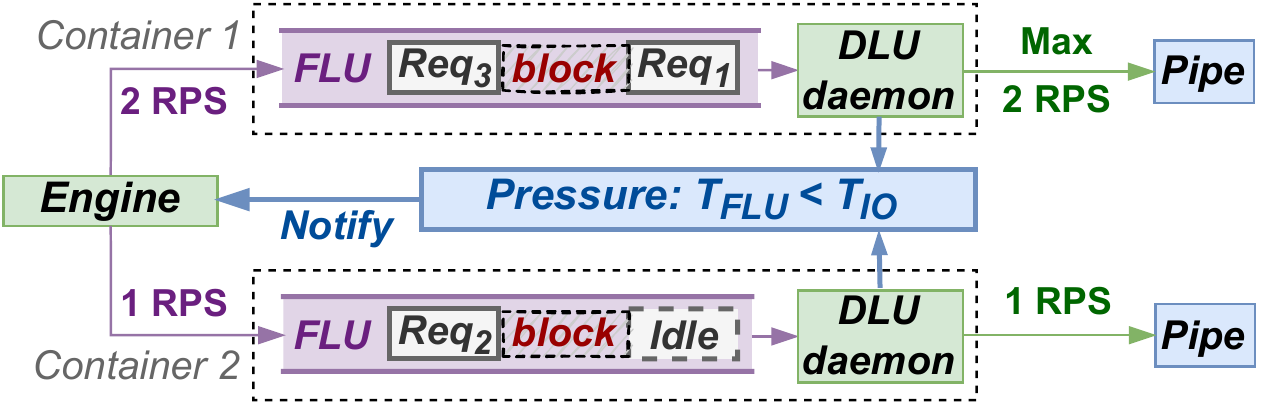}}
    \caption{\label{fig:pressure} The pressure-aware function scaling mechanism.}
\end{figure}

With the FLU-DLU abstraction, the data transfer of a function becomes an asynchronous operation. 
The asynchronous operation may result in a long response latency.
Figure~\ref{fig:pressure}(a) shows such an example. In the figure, the FLU receives 3 function requests per second, but the DLU daemon is only able to transfer the data to the destination 
for 2 requests per second. 
In this case, the queuing effect can result in severe long response latency throughout the serverless workflow.

It is not a problem if the data transfer is a synchronous operation, as in the control flow paradigm, because the serverless engine will create new containers for subsequent invocations when the container is processing a function.
The situation is even worse for small function containers (e.g. 128MB), which are allocated with a lower upper limit of network bandwidth. 
A technique is required to tackle the problem due to asynchronous data transfer with the DLU daemon. 

We therefore propose a pressure-aware function scaling mechanism as shown in Figure~\ref{fig:pressure}(b). 
Specifically, before the DLU receives data from the FLU, it will simultaneously acquire the average execution time $T_{FLU}$ and the size of the data to be transferred $Size$ of this FLU computation. On the basis of a prior knowledge of the container's network bandwidth $B_w$, Equation (\ref{eq:pressure}) calculates the expected data transfer pressure.
\begin{equation}
  \label{eq:pressure}
  \small
  Pressure(FLU_f) = \alpha\frac{Size}{B_w}-T_{FLU}
\end{equation}

In the equation, $Size/B_w$ represents the ideal transfer time, and $\alpha$ indicates the loss factor during actual data transfer. This loss factor is determined by the characteristics of the connector implementation.
Suppose $ Pressure(FLU_f)\le 0 $, the DLU can entirely consume the data generated from the FLU computations and does not need to scale up the containers. In this case, queries should remain dispatched simply based on idle FLUs within containers. 
Otherwise, if $ Pressure(FLU_f)>0$, backpressure happens, and the DLU sends a Callstack blocking signal to the workflow engine to block its FLU for $ Pressure(FLU_f)$ time.

In this way, the FLU of the container cannot serve subsequent invocations, and the FLU-producing rate is limited to the maximum DLU-consuming efficiency. 
At the same time, the engine follows the serverless manner to scale out the containers to handle the remaining workflow invocations.

\section{Serverless Workflow Expression by Data-flow}
\label{sec:engine}
\begin{figure}
  \centerline{\includegraphics[width=.84\columnwidth]{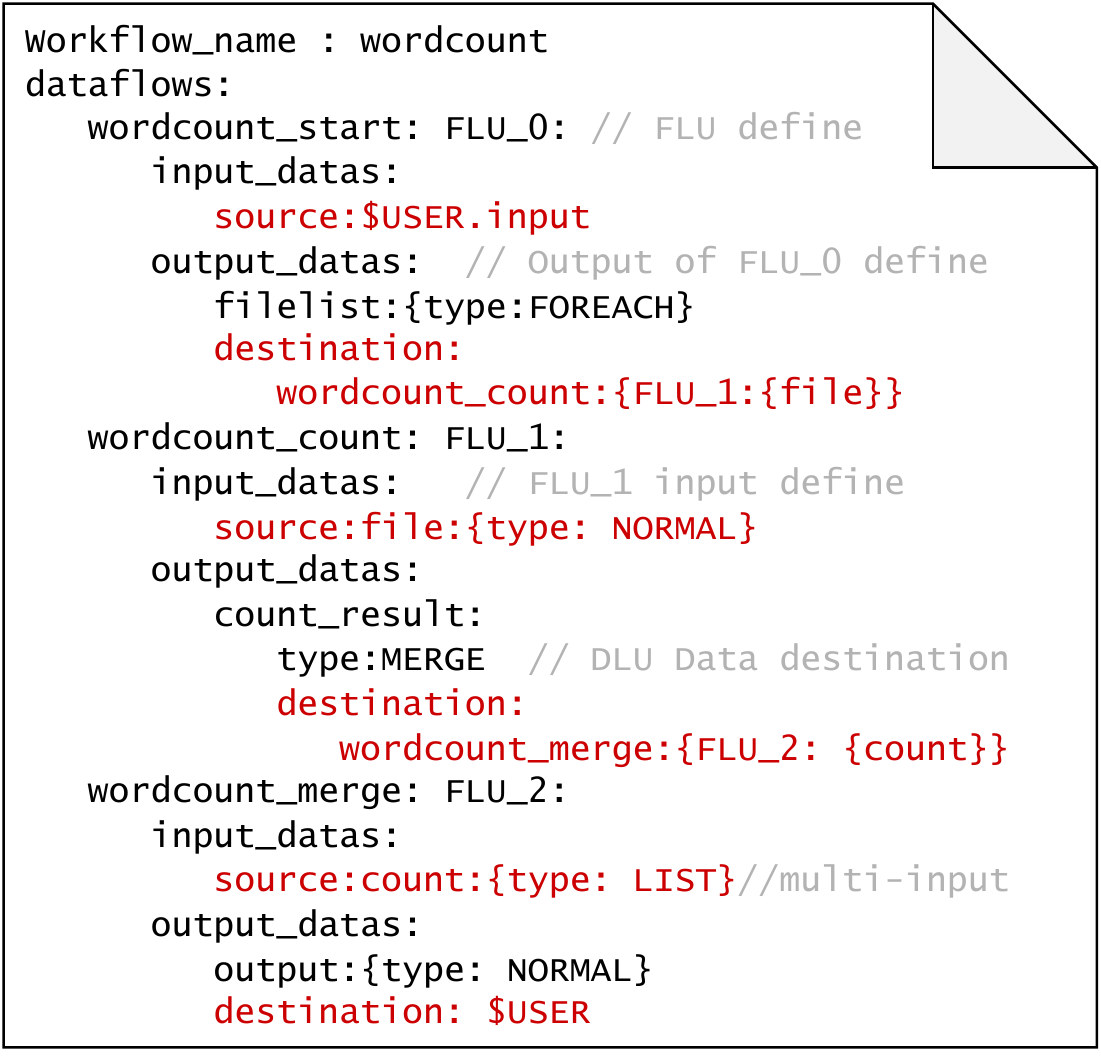}}
  \caption{\label{fig:yaml} Pseudocode of defining the data transfer relationship of functions in the benchmark {\it WordCount}.}
\end{figure}

With the FLU and DLU abstraction, the computation and the communication is decoupled. 
However, the data-flow graph for a workflow is still unavailable only with the abstraction, as the destination of a function's data is still not defined. With control-flow, the programmers need to define the function triggering relationships, the input data, etc. Similarly, the programmers need to define the output data and the destination of a function in DataFlower.
As an example, Figure~\ref{fig:yaml} shows the pseudocode of defining the data transfer relationship for the benchmark {\it WordCount}.
In general, for each FLU, we need to define the source of the inputs and the destination of the outputs.
However, a serverless workflow is still not able to run only with the data transfer relationship. 
This is because a function does not know on which node and containers the destination function of its outputs runs.

\subsection{Serverless Workflow Scheduling with DataFlower}
\begin{figure}
  \centerline{\includegraphics[width=.92\columnwidth]{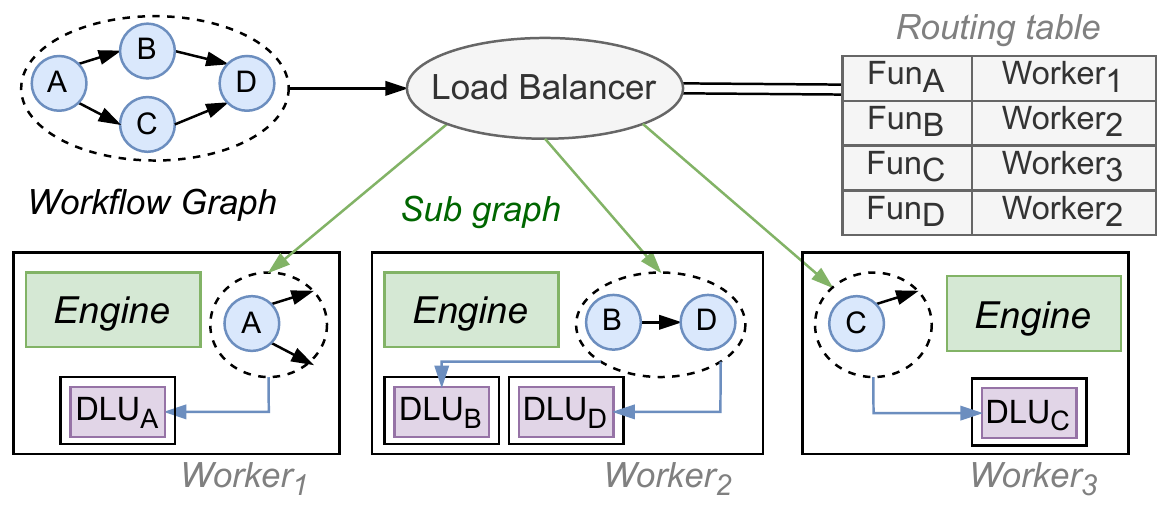}}
  \caption{\label{fig:loadbalancer} The upper load balancer maps functions to each worker node for data-flow graph parsing.}
\end{figure}

To resolve the above problem, DataFlower obtains the function deployment information from the load balancer. 
Figure~\ref{fig:loadbalancer} shows the way to parse the data-flow graph. 
In order to minimize the overhead of accessing the data-flow graph, the workflow scheduling engine in DataFlower is designed to be a decentralized component.
As shown in the figure, DataFlower deploys a workflow scheduling engine on each node that hosts at least one function of the serverless workflow.
Each engine parses the data-flow graph based on the code in Figure~\ref{fig:yaml} and the function deployment information. 
For an engine, only the part of the data-flow graph related to the functions on the local node needs to be parsed.



The engine maintains a pool of containers for local functions.
When a function is triggered, the local workflow engine allocates an idle container FLU or cold starts a new container to run it.
The auto-scaling policy in serverless computing determines whether to scale out a new container for a function. In addition, the pressure-aware function scaling mechanism in Section~\ref{sec:pressure_scaling} can also suggest starting a new container for a function due to queuing on the DLU.

Note that DataFlower does not rely on a specific load balancer.
DataFlower exposes an interface to the upper load balancer for customized function deployment policies. 
DataFlower provides explicit control of the data-plane, allowing it to work with various load balancing strategies for high extensibility and flexibility.

\subsection{Fault-tolerance Model and Data Consistency}


DataFlower's fault tolerance model supports at-least-once or exactly-once execution semantics based on data flow persistence and integrity. 
The scheduling engine ensures that a function is not triggered if its predecessor fails or only sends partial data due to the data plane interrupt. 
In addition, the pipe connector itself periodically creates checkpoints asynchronously and incrementally to handle error retries. Based on the checkpoints,
the workflow scheduling engine locates the workflow invocation to be retried, identifies the last pipe connector with correct data checkpoints, 
and \textit{ReDo} the associated failed function execution through backtracking.

In addition, DataFlower's workflow scheduling engine uses a customized container keep-alive strategy to ensure data consistency. 
With the control flow paradigm, a container can be safely recycled if it is not processing a function. 
With the data flow paradigm, a container FLU may not be processing a function, but some destination data sinks may still be pumping new data from the container DLU through a pipe connector. 
In this case, if the container is identified as idle and recycled, data consistency will not be guaranteed.
To this end, when using a keep-alive strategy, DataFlower ensures that a container is not recycled unless 
the container FLU is idle and there is no data remaining in the container DLU to be pumped.

\section{Host-Container Collaborative Communication Mechanism}
\label{sec:communication}

\begin{figure}
  \centerline{\includegraphics[width=.87\columnwidth]{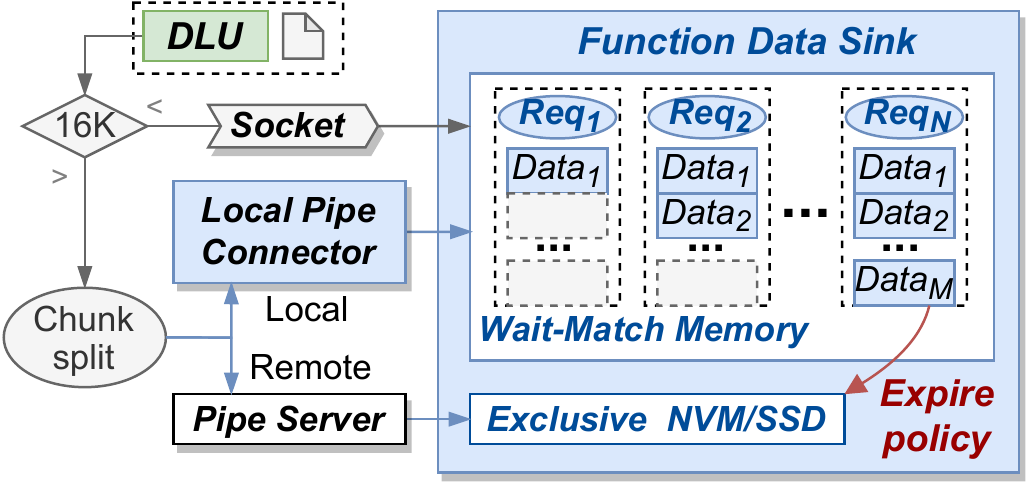}}
  \caption{\label{fig:sec3-3} The function locality-aware pipe connector establishing, and the data structure organization in the data sink.}
\end{figure}

With the data-flow paradigm, a function is triggered to run based on the data availability. 
A container will not be assigned until all input data of a function is ready, which indicates that the input data cannot be loaded immediately by the corresponding FLU.
To cache and manage the input data for a function before triggered, we propose a host-container collaborative communication mechanism, as shown in Figure~\ref{fig:sec3-3}.

Specifically, DataFlower uses a {\it data sink} on each host node to temporarily 
cache the input data of all the functions on the node. 
When the DLU of a function determines to transfer the data to its destination function, 
a pipe connector is built between the source node and the destination node. 
If a single predecessor has to communicate with many child functions, The DLU of the predecessor will send the data to child functions through different pipe connectors in a FIFO fashion. 
The data is transferred to the data sink of the host node that deploys the destination function.
Once the function is triggered, the corresponding container fetches the data from the data sink.



There are two ways to implement the pipe connector. 
A local pipe connector is established when the data-plane in the DLU indicates that two functions are on the same node. 
In this case, the data stream is pumped directly to the data sink. 
If the two functions are on different nodes, a pipe connector that supports data streaming is used to transfer the data across the nodes.
For small data under 16K, DLU does not use a pipe connector and splits it, passing it directly to the destination via socket. Such a pipe connector mechanism can fully exploit function locality to improve data-flow efficiency.

It is also possible for a function to require input from multiple functions (e.g. merge), and the input from these functions may arrive in different order in different invocations of the serverless workflow.
To boost the data loading of the DLU, we allocate a dedicated Wait-Match Memory~\cite{DBLP:journals/tse/HerathYSY88} to maintain the cached inputs of a function. The Wait-Match Memory acts as a high-performance key-value store, and we design a multi-level index structure (\textit{RequestID}, \textit{FunctionName}, \textit{DataName}) to further reduce the indexing overhead of a huge data sink.

Another challenging problem is that a function data sink maintains different data from different \textit{RequestID} and cannot consume the data and trigger the FLUs in time. As shown in Figure~\ref{fig:sec3-3}, when the source function first pumps $Data_M$ of $Req_N$ into the data sink, it will continue to occupy memory until the remaining data copies of $Req_M$ arrive. 
To minimize the memory overhead for data caching, 
DataFlower proposes a hybrid mechanism to perform fine-grained data management in the Wait-Match memory and alleviate the data backlog. 
\begin{itemize}
  \item \textbf{Proactive release.} 
  As soon as all destination FLUs have received the data, this intermediate data can be proactively released from the Wait-Match Memory. In the control-flow paradigm without prior knowledge of data dependencies, the caching design such as FaaSFlow can only remove the cache after each request completion.
  \item \textbf{Passive expire.} To further reduce the data footprint in Wait-Match memory, we use the passive expire mechanism based on a cache timeout. In this method, each piece of data is given a TTL (Time To Live), and it will expire and persist to the function-exclusive disk once timeout occurs. 
\end{itemize}

The proactive release approach removes useless data, and the passive expire approach guarantees the data freshness in the memory cache. Therefore, both can increase the cache hit rate with limited Wait-Match memory. 


\section{Implementation of DataFlower}
\label{sec:implementation}
We have implemented and open-sourced DataFlower, which involves four modules: decentralized workflow engine (1466 LOC), function manager/invoker (546 LOC), and container encapsulation (902 LOC),  excluding third-party libraries. In our implementation, each FLU is a thread created by the executor in the container, and the \textit{WORKDIR} of each thread is exported into the container as an environment variable via the hash value of the user \textit{RequestID}. We use a fixed 15-min keep-alive strategy for each FLU in the container. The computing resources of each function container are limited by Linux Cgroup, and the network is also limited by Linux TC.

Considering the performance degradation of the document-oriented CouchDB~\cite{CouchDB} via REST APIs, 
we implement the pipe connector based on kafka~\cite{kafka} that supports event streaming for high-performance data pipelines. 
In the Kafka-based pipe connector, each data-flow corresponds to a topic, and each container is assigned a partition by default. 
The data sink implementation is based on a local SSD bound to the volume (for inter-node data storage) and Redis~\cite{redis} (for intra-node data cache). The fault tolerance of maintaining states in Wait-Match Memory and exclusive NVM/SSD depends on the basic fault tolerance mechanism that provided by Kafka and Redis. The communication within the node consists of two aspects, the triggering of function requests and the data-plane synchronization during the execution of the DataFlower. The function invocations are triggered by REST APIs with the executor inside the container, while the interaction between the executor, the DLU daemon, and the host workflow engine establishes socket-based connections. DataFlower also uses a CouchDB server to collect the logs during execution. 


\section{Experimental Evaluation}
\label{sec:eval}
In this section, we first evaluate DataFlower in reducing the response latency and increasing the peak throughput. 
Then, we break down the effectiveness of the mechanisms in DataFlower and show the adaptiveness of DataFlower for various workflows.
Lastly, we also discuss the architectural implications.

\subsection{Experimental Setup}
\label{sec:setup}

We evaluate DataFlower on a 5-node cluster. We use one node to generate workflow invocations, one node to be the backend storage node, and three nodes as the worker nodes. CouchDB~\cite{CouchDB}, a widely-used document-oriented database, is used to be the backend storage for persisting the intermediate data (replaced with one Kafka node to support pipe connector for DataFlower). 
Each node is equipped with Intel Xeon (Ice Lake) Platinum @3.5GHz CPU (the load generator node and the backend storage node have 8 cores and 16GB memory, and 
each worker node has 16 cores and 64G memory) with 200GB SSD (3000 IOPS). 
We still use the four best practice serverless workflows, \textit{Video-FFmpeg} ({\it vid})~\cite{ffmpeg}, \textit{ML-based Image Processing} ({\it img})~\cite{imagemagick}, \textit{Singular Value Decomposition} ({\it svd})~\cite{DBLP:journals/pomacs/MahgoubYSMEBC22} and \textit{WordCount} ({\it wc})~\cite{DBLP:conf/icde/ZhangHDZH17}, as benchmarks.



The container used to run a function is configured with the practical specification.
Based on the previous study~\cite{DBLP:conf/nsdi/PuVS19}, the container network bandwidth increases as the container scales up. Following this pattern, we allocate 0.1 core and 40Mbps network bandwidth for a 128MB-sized container.
The resources are allocated proportionally according to the container memory size.
We also evaluate the impact of the container specification on the performance of DataFlower in Section~\ref{sec:resource-scal}.



We compare DataFlower with sota serverless workflow systems, FaaSFlow~\cite{DBLP:conf/asplos/LiLGCCZG22} and SONIC~\cite{DBLP:conf/usenix/MahgoubSMKCB21}. In FaaSFlow, a decentralized scheduling pattern is used to reduce the scheduling overhead in the workflow, and enable data transferring through memory for co-located functions. 
We implement SONIC by replacing the backend storage in FaaSFlow with local storage. The data to be transferred is persisted in the host, and then each destination function container builds a peer-to-peer connection with the source storage to fetch data in parallel.


We evaluate DataFlower with both synchronous invocations and asynchronous invocations~\cite{asynandsyn}, as 
production serverless systems provide the two patterns.
With the synchronous invocation pattern, the requests are generated in a closed-loop, where a new request is generated after the previous request completes. We increase the load in this scenario by increasing the number of client threads that generate the requests.
With the asynchronous invocation pattern, the requests are generated in an open-loop with a given load.
The results with the synchronous invocation and asynchronous invocation patterns reveal the peak throughput that can be achieved with these systems, and the tail latency at a given load, respectively.

\begin{figure*}
  \captionsetup[subfloat]{margin=6pt,format=hang}
  \vspace{-3mm}
  \centering
  \subfloat[Asynchronous $img$]{\includegraphics[width=.41\textwidth]{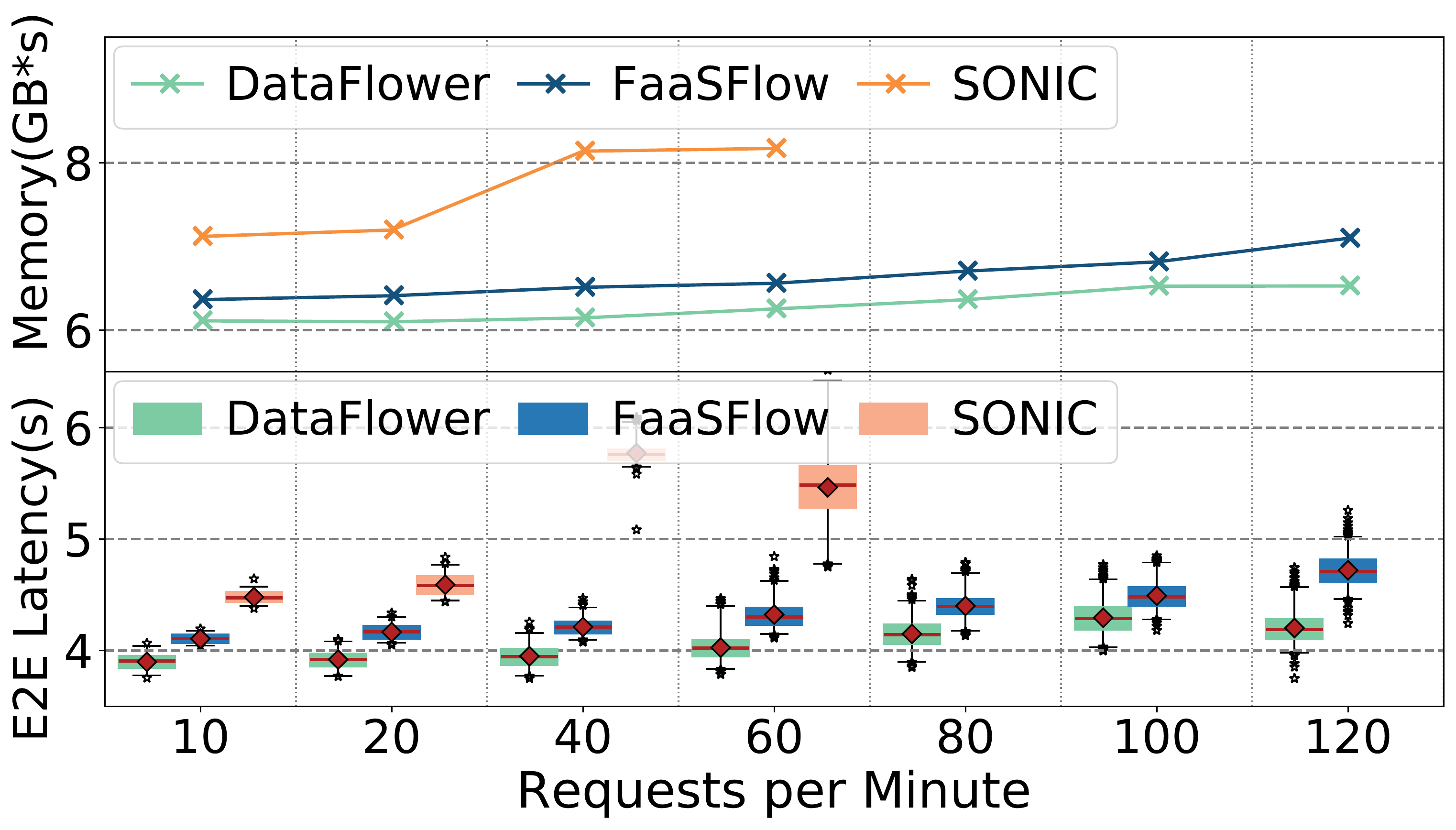}} \hspace{8mm}
  \subfloat[Asynchronous $vid$]{\includegraphics[width=.42\textwidth]{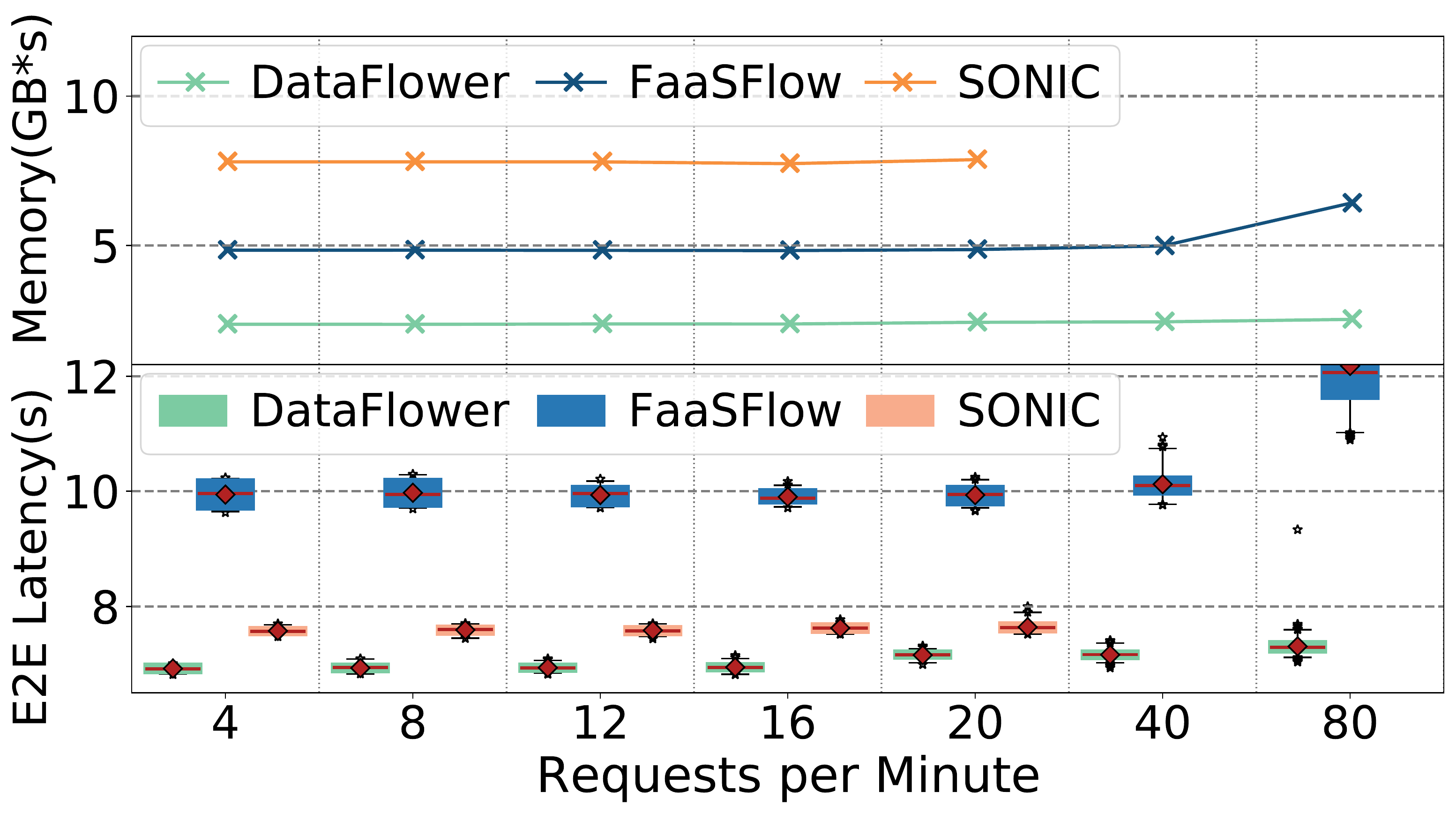}}
  \\ \vspace{-3mm}
  \subfloat[Asynchronous $svd$]{\includegraphics[width=.419\textwidth]{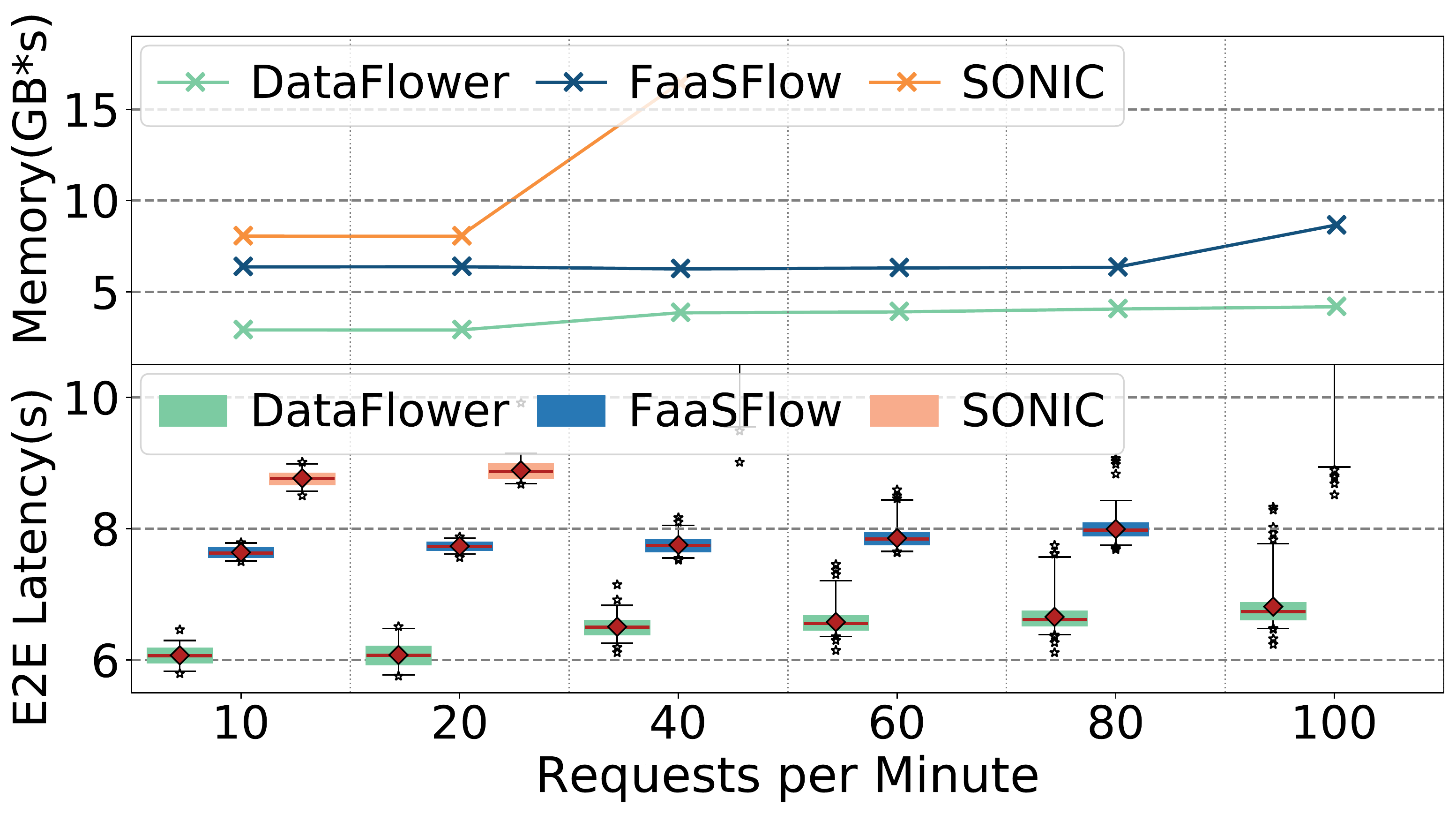}} \hspace{7mm}
  \subfloat[Asynchronous $wc$]{\includegraphics[width=.425\textwidth]{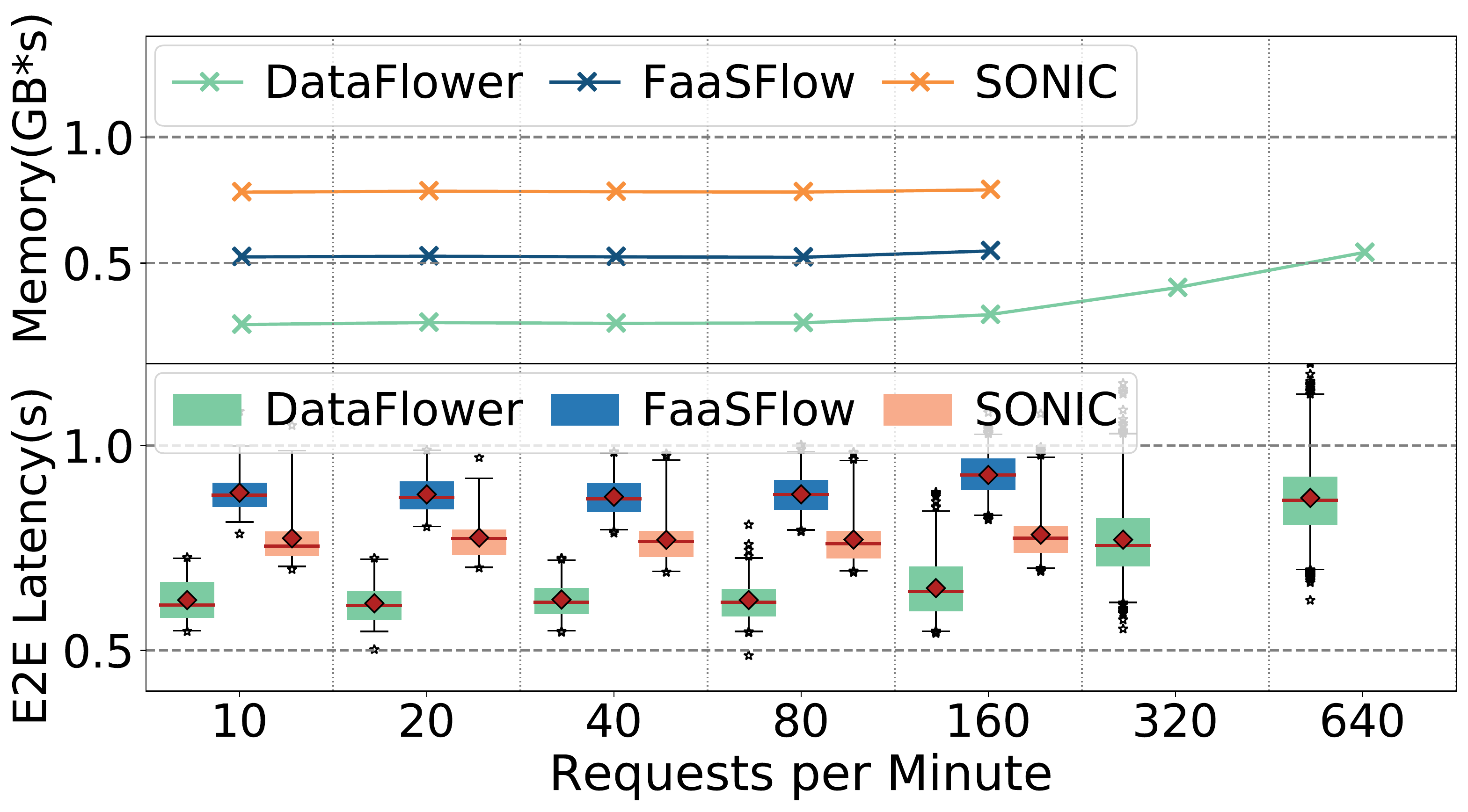}}
  \hspace{1mm}
  \caption{\label{fig:sec5-2-1} The end-to-end latency and memory usage of the benchmarks with the asynchronous load generation pattern.}
\end{figure*}

\begin{figure*}
  \vspace{-2mm}
  \captionsetup[subfloat]{margin=4pt,format=hang}
  \centering
  \subfloat[Synchronous $img$]{\includegraphics[width=.267\textwidth]{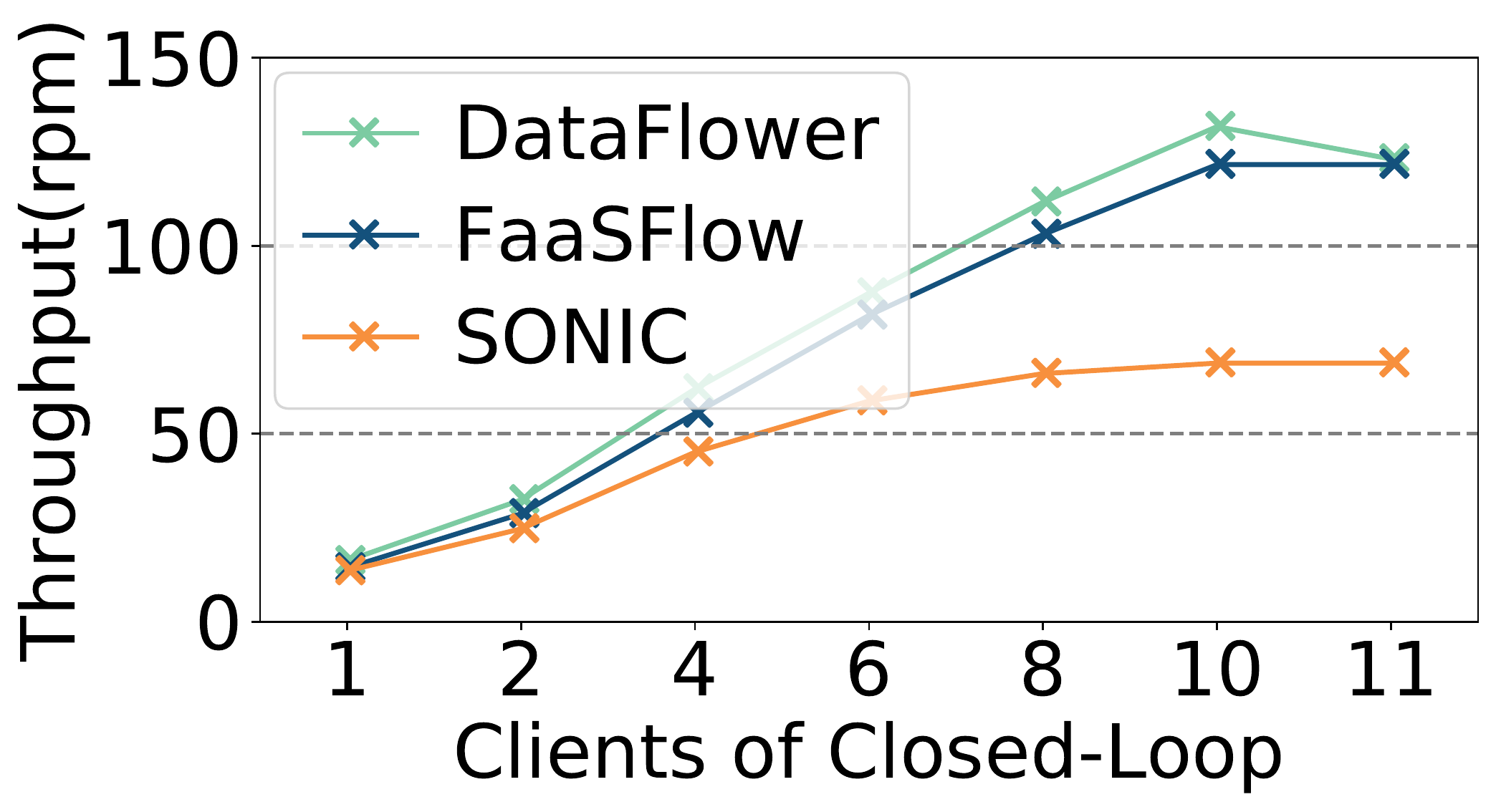}}
  \subfloat[Synchronous $vid$]{\includegraphics[width=.238\textwidth]{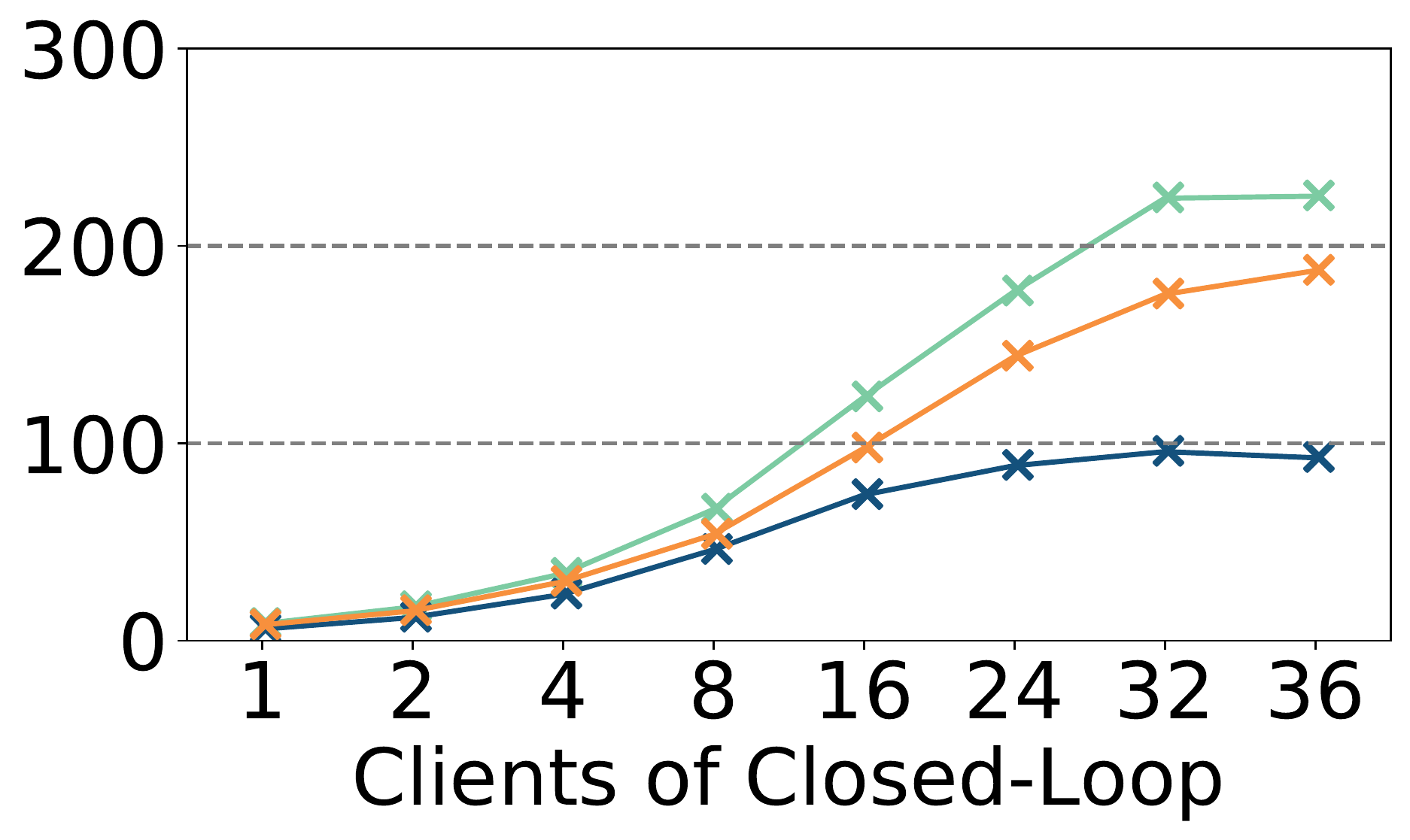}}
  \subfloat[Synchronous $svd$]{\includegraphics[width=.238\textwidth]{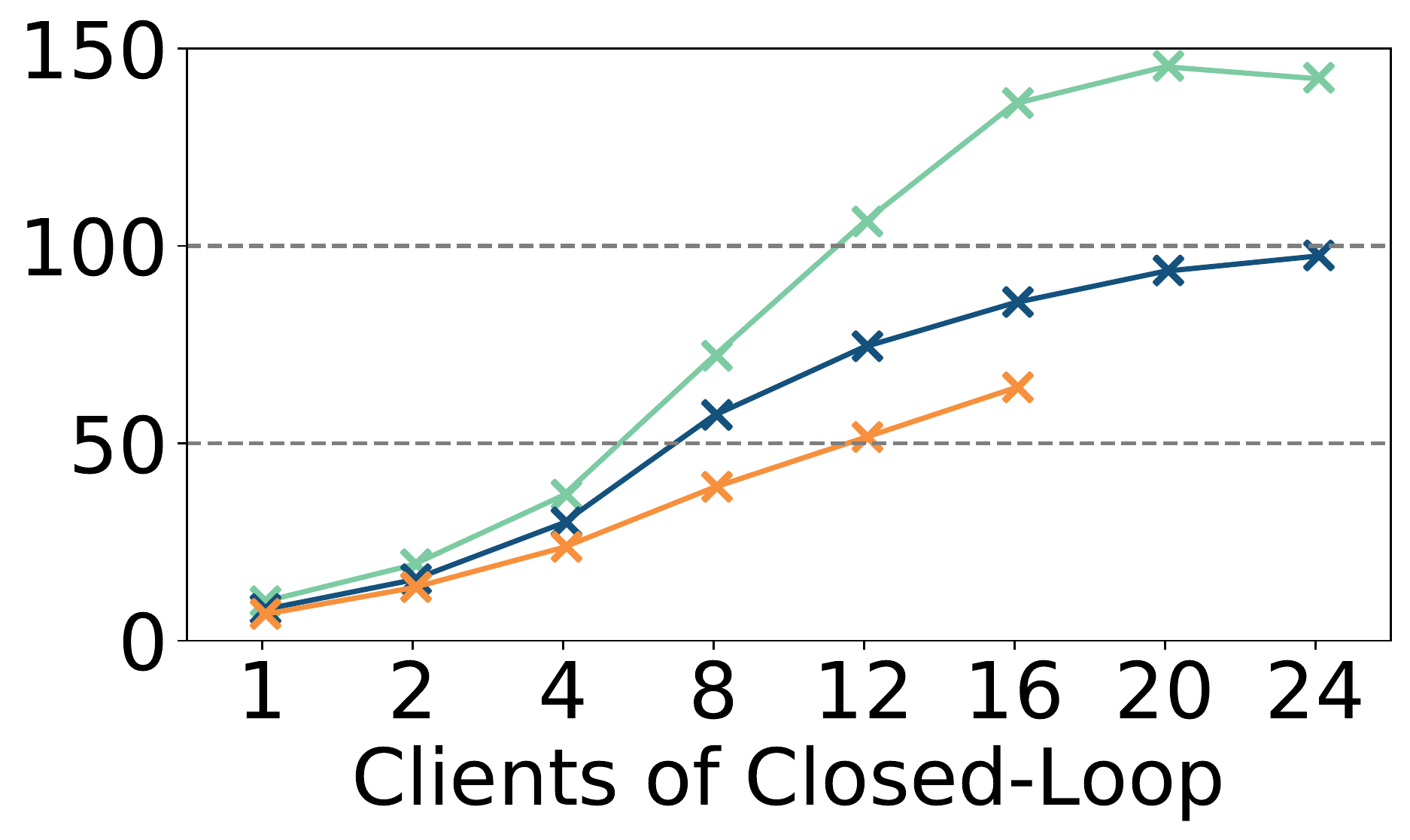}}
  \subfloat[Synchronous $wc$]{\includegraphics[width=.248\textwidth]{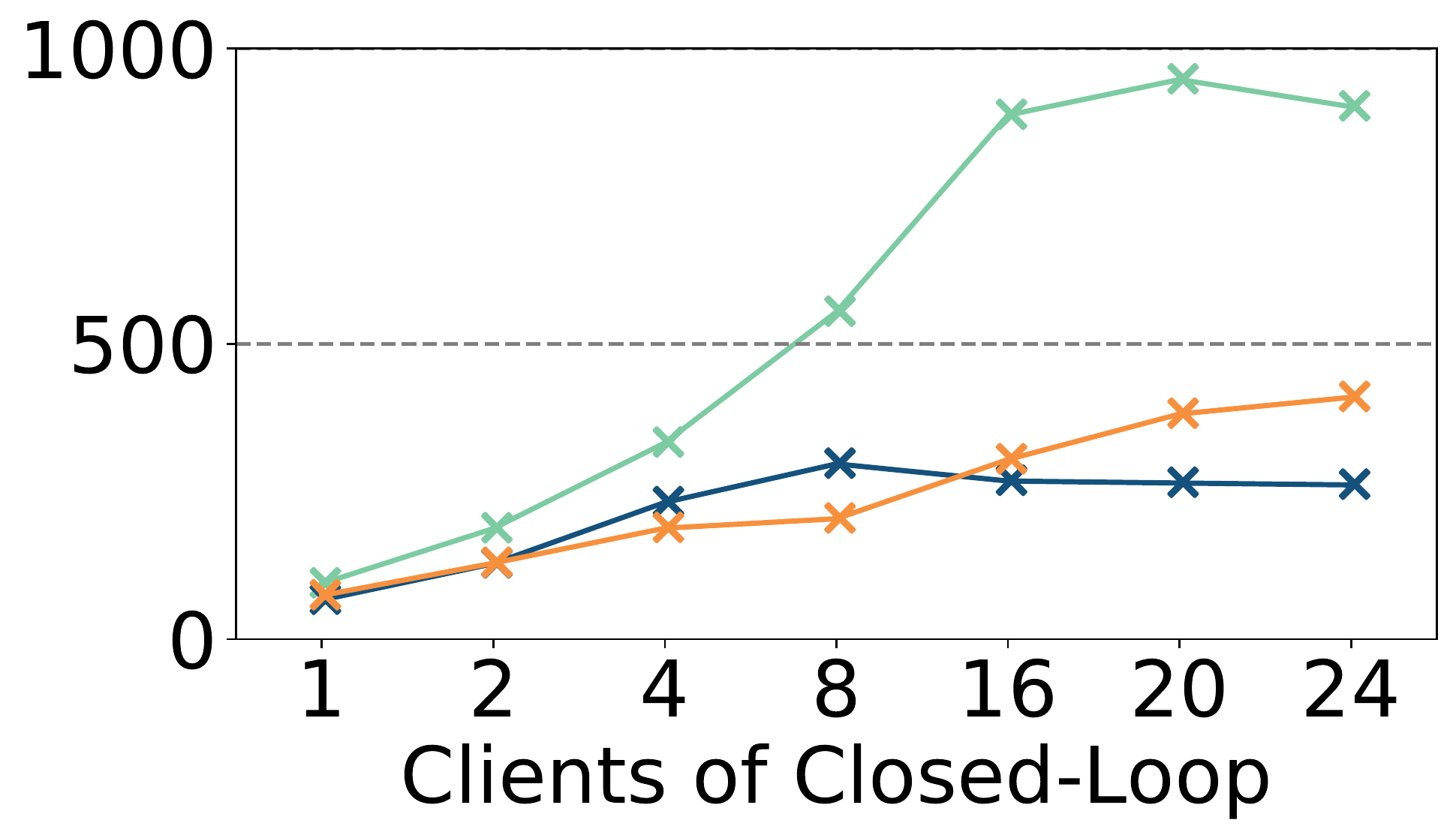}}
   \caption{\label{fig:sec5-2-2} The throughput of the benchmarks when different numbers of clients generate requests synchronously.}
\end{figure*}


\subsection{Response Latency and Peak Throughput}
\label{sec:evalsyn}




We first report the result with the asynchronous invocation pattern.
Figure~\ref{fig:sec5-2-1} shows the end-to-end response latencies and the memory resource usage of the benchmarks at different loads.
Suppose $N$ GB memory is occupied by $t$ seconds, and the memory resource usage is calculated to be $N\times t$. 
The metric quantifies the cost of running the serverless workflow.
The missing points/boxplots in the figure mean the benchmark suffers from timeout due to the huge latency. 





DataFlower reduces both the average and tail latency of the benchmarks compared with FaaSFlow and SONIC.  
Specifically, it reduces the 99\%-ile latency of the benchmarks by 5.7\% to 35.4\% compared with FaaSFlow, and by 8.9\% to 29.2\% compared with SONIC.
DataFlower reduces the response latency, because the input data of the function is already prepared on the host. In this way, DataFlower does not need to fetch the data from the backend storage during the function execution as FaaSFlow does.
SONIC also directly obtains the data from the source function to its local storage. 
However, the limited bandwidth of each container results in a long data transfer time. The slow data transfer further incurs the queuing of requests at the containers at high load, without the pressure-aware scaling mechanism in DataFlower.

We can also observe that memory usage reduction is even higher than the reduction of response latency. DataFlower reduces container memory usage by 19.1\% to 69.3\%, and by 7.4\% to 64.1\% compared with FaaSFlow and SONIC, respectively.
This is mainly because the response latency reduction may have multiple effects on memory reduction for parallel processing, such as map-reduce logic. Each branch container can benefit from memory usage reduction, though the critical execution path can only reflect the end-to-end tail latency.



Figure~\ref{fig:sec5-2-2} shows the serving throughput (requests-per-minute, rpm) of benchmarks when different numbers of clients generate requests synchronously.  
As shown, DataFlower increases the achievable peak throughputs of the benchmarks by 1.03X to 3.8X compared with FaaSFlow,
and by 1.29X to 2.42X, compared with SONIC.
The throughput is saturated as either CPU or network bandwidth becomes the bottleneck when the number of clients increases.
We can also observe that {\it svd} fails with SONIC when there are 20 or more load generation clients in the closed-loops. This is because the application-aware data passing in SONIC can only optimize the data transfer of a single workflow invocation, but cannot recognize the transmission bottleneck of scaling containers for parallel workflow invocations.

DataFlower increases the peak throughputs of serverless workflows because its FLU-DLU abstraction enables the computation-communication overlap. 
\subsection{Effectiveness of Pressure-aware Function Scaling}
\label{sec:evalasyn}

\begin{figure*}
  \centering
  \vspace{-4mm}
  \subfloat[Pressure-aware for $img$]{\includegraphics[width=.267\textwidth]{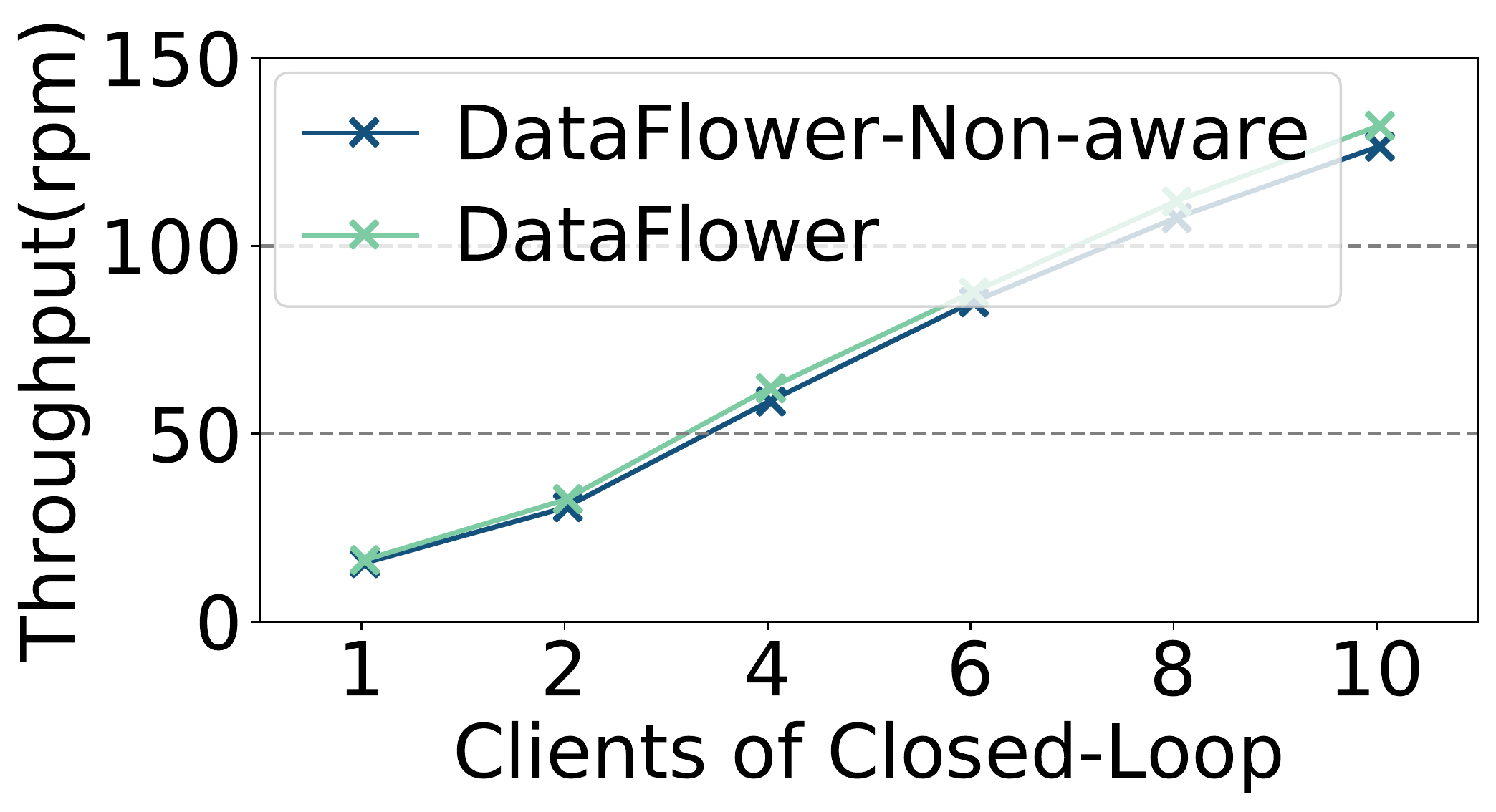}}
  \subfloat[Pressure-aware for $vid$]{\includegraphics[width=.238\textwidth]{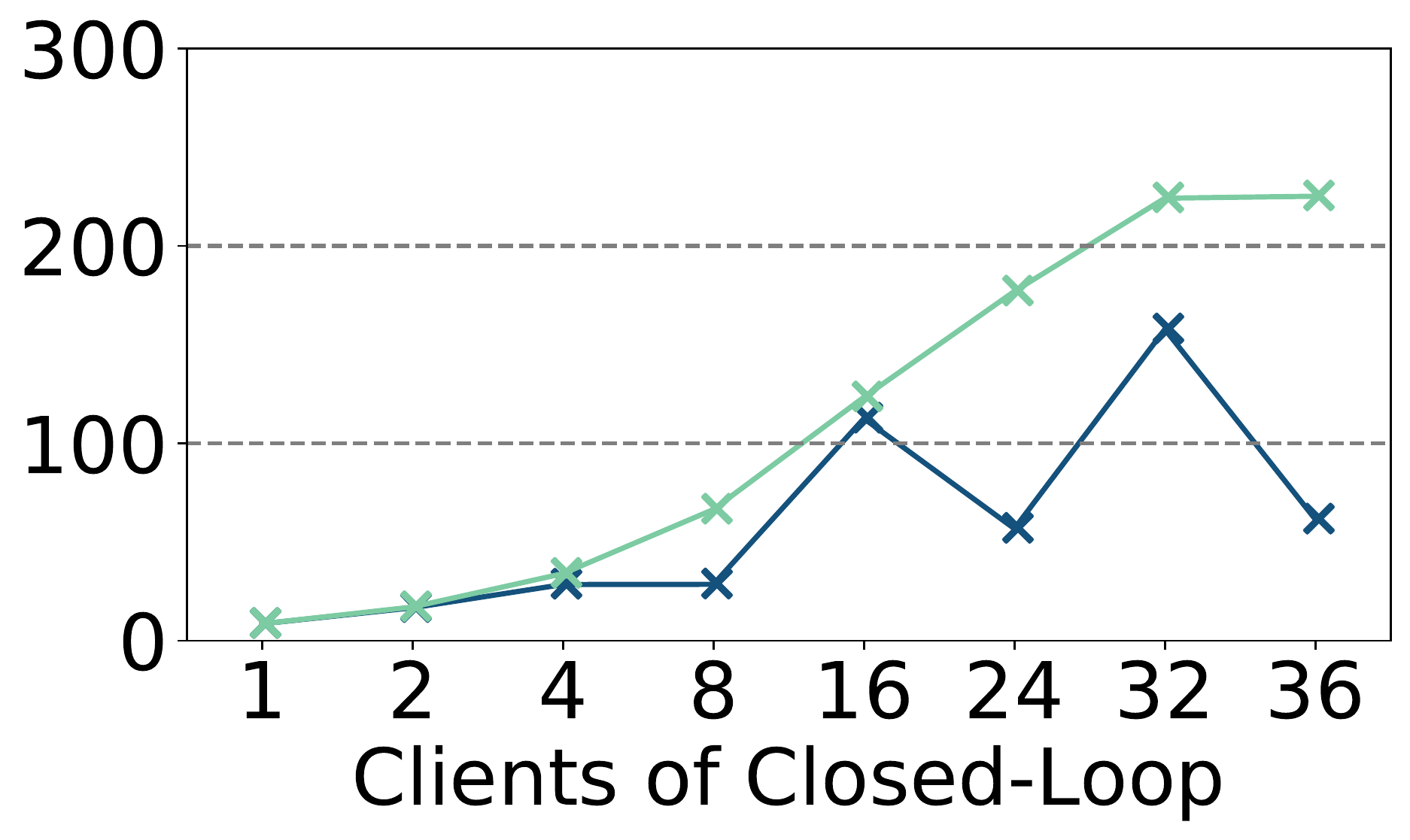}}
  \subfloat[Pressure-aware for $svd$]{\includegraphics[width=.238\textwidth]{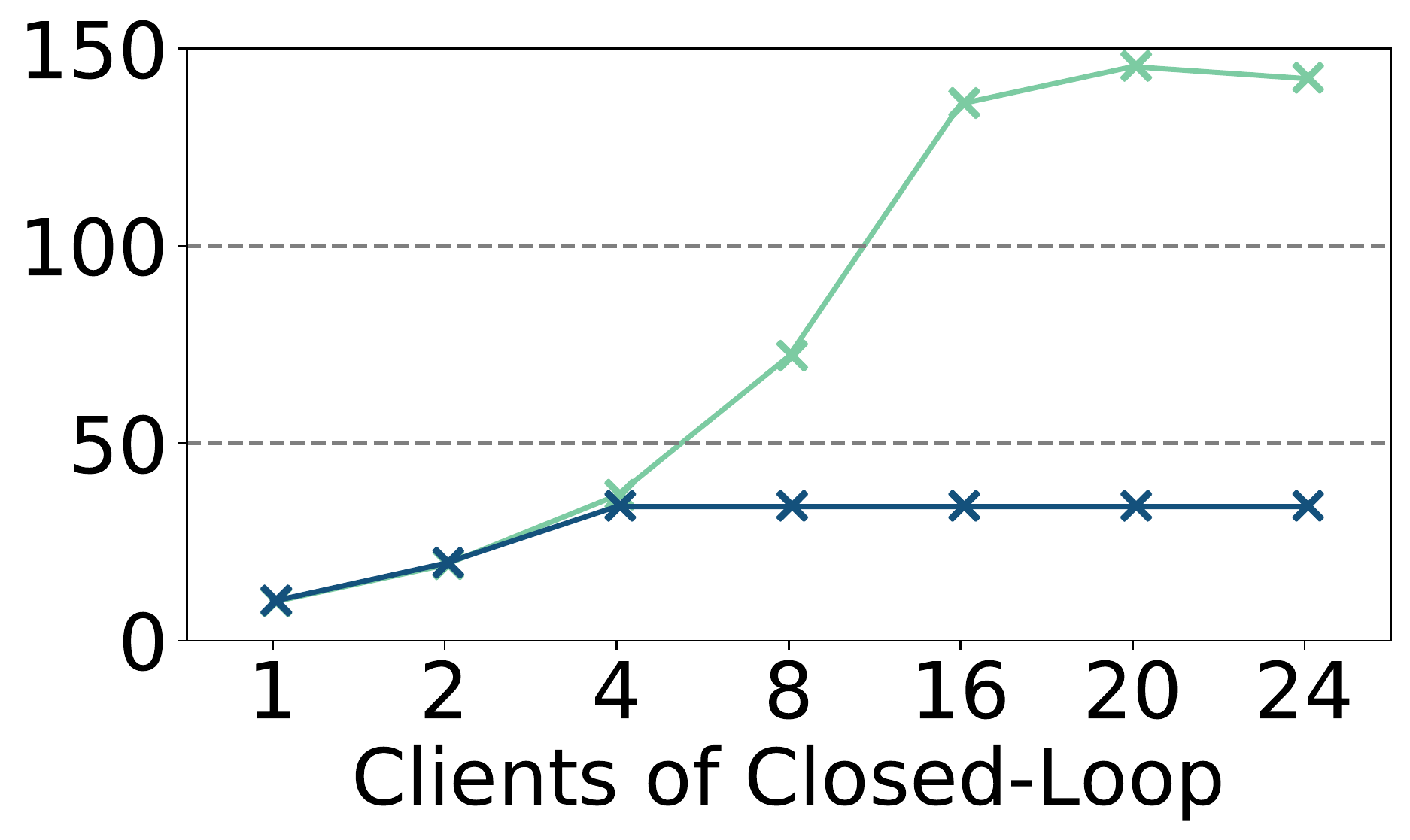}}
  \subfloat[Pressure-aware for $wc$]{\includegraphics[width=.248\textwidth]{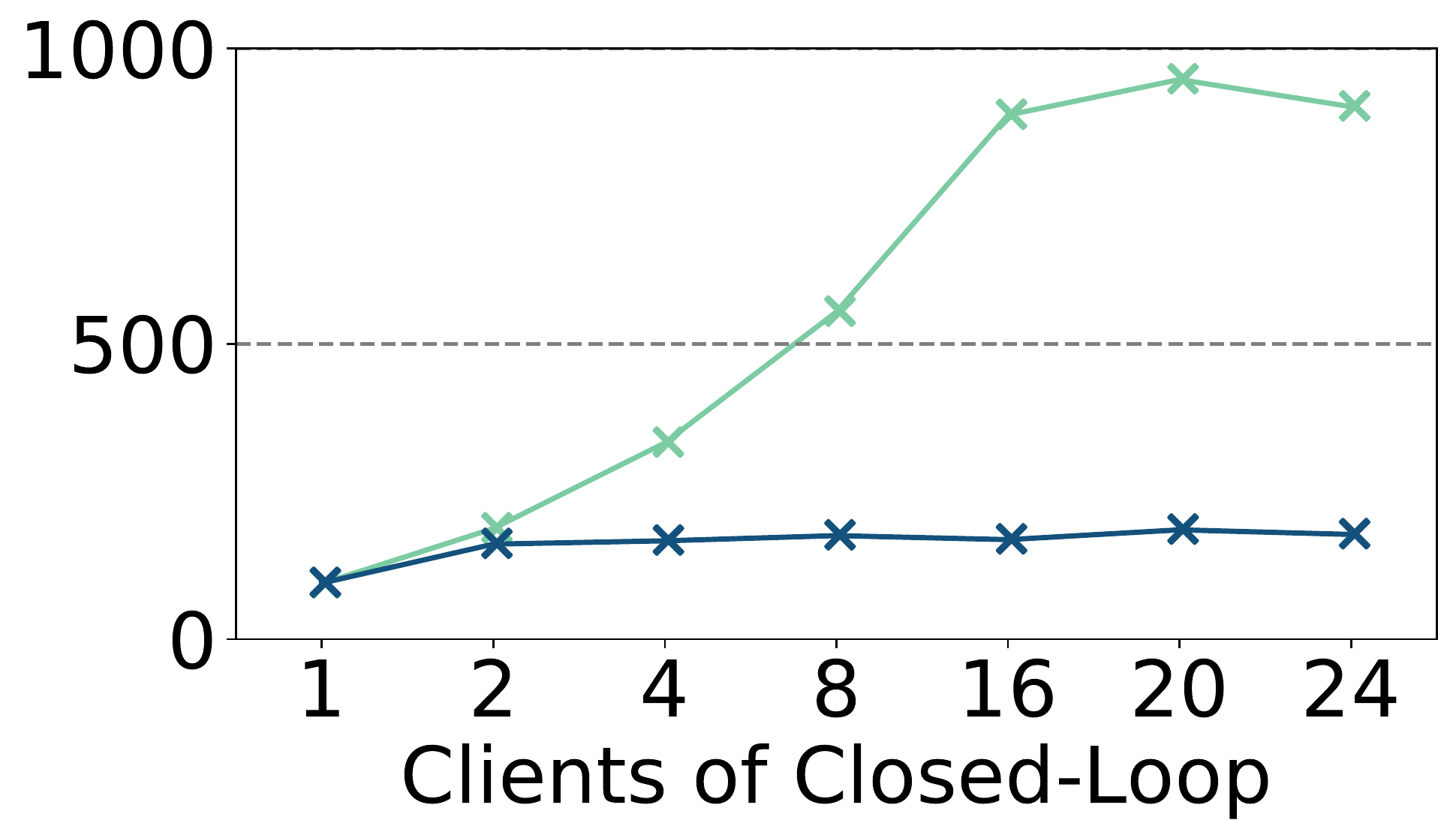}}
    \caption{\label{fig:pressure_aware} The achieved throughput of the benchmarks in DataFlower and DataFlower-Non-aware.}
    \vspace{2mm}
\end{figure*}

In this experiment, we compare DataFlower with DataFlower-Non-aware, a variant of DataFlower that 
disables the pressure-aware function scaling.
Figure~\ref{fig:pressure_aware} shows their performance with the synchronous load pattern.
The achieved throughput drops significantly with DataFlower-Non-aware.

In the four benchmarks, DataFlower and DataFlower-Non-aware perform similar for {\it img}. 
This is mainly because the intermediate data between functions in {\it img} is small. A DLU can always transfer the output of a function to the destination without obvious queuing. 
On the contrary, for data-intensive functions in {\it vid}, {\it svd} and {\it wc}, the data transfer time of DLU is much longer than the computation time of FLU. 
Without the pressure-aware function scaling in DataFlower-Non-aware, 
user requests queue at DLUs. The peak throughputs of these benchmarks are constrained by the data transfer bottleneck.

We can also observe that DataFlower-Non-aware achieves relatively high throughput when there are 16 and 32 closed-loop synchronous load clients for {\it vid}.
This is because FLUs in current containers are insufficient to serve the increasing workload, and the serverless platform automatically scales out containers. This partially alleviates the queuing on the DLUs.

The pressure-aware function scaling is able to notice the queuing on the DLUs and scale out the containers, even if the containers are enough in terms of computation ability.

\subsection{Effectiveness of Early Triggering and Input Caching}
\label{sec:early}
In this experiment, in order to eliminate the impact of network communication on the function triggering, we force all functions of a benchmark to run on a single node. 
In this way, both DataFlower and FaaSFlow communicate via  local memory.

\begin{figure}
  \centerline{\includegraphics[width=\columnwidth]{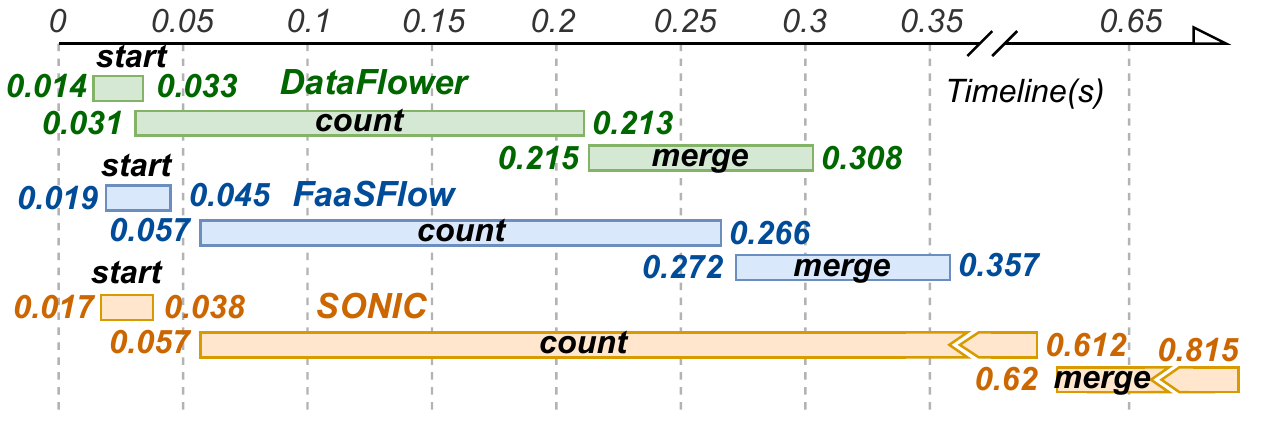}}
    \caption{\label{fig:timeline} The function triggering timeline of benchmark {\it wc} with DataFlower, FaaSFlow, and SONIC.}
\end{figure}

Figure~\ref{fig:timeline} shows the execution timeline of the functions {\it start}, {\it count}, and {\it merge} in the benchmark {\it wc}. 
Other benchmarks show a similar result. 
The function {\it count} is triggered before the function {\it start} completes, and {\it merge} is triggered 2$ms$ later once {\it count} completes with DataFlower.
On the contrary, {\it count} and {\it merge} are triggered 15$ms$ and 6$ms$ later than their predecessor completion in FaaSFlow.
The late triggering is due to the control-flow paradigm, though the 
intermediate data is also transferred through the local memory with FaaSFlow.
On the other hand, the functions are triggered much later with SONIC, because the functions state still communicate through local VM storage rather than shared memory.
\begin{figure}
  \captionsetup[subfloat]{margin=2pt,format=hang}
  \centering
  \subfloat[$img$]{\includegraphics[width=.215\textwidth]{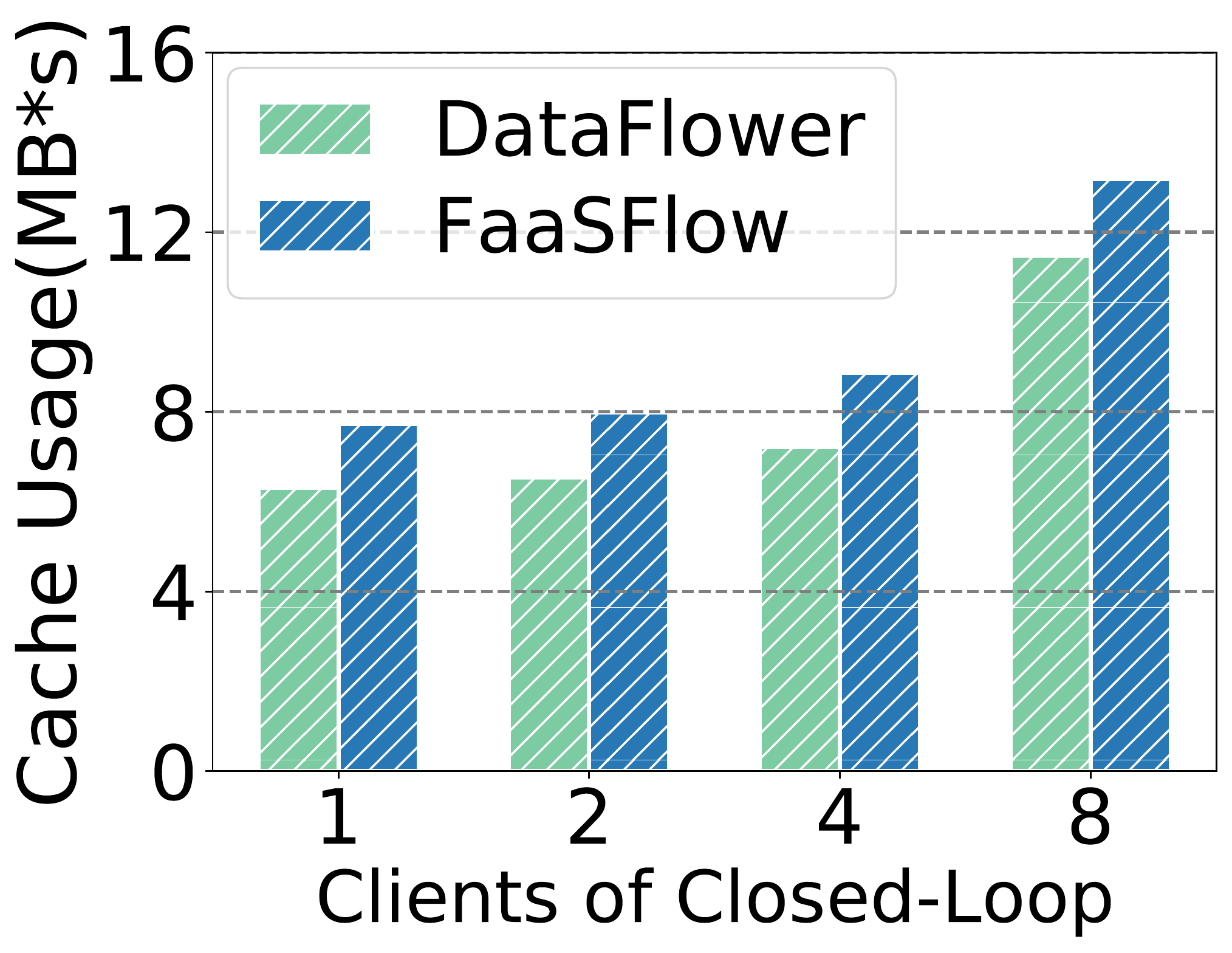}} \hspace{1mm}
  \subfloat[$vid$]{\includegraphics[width=.18\textwidth]{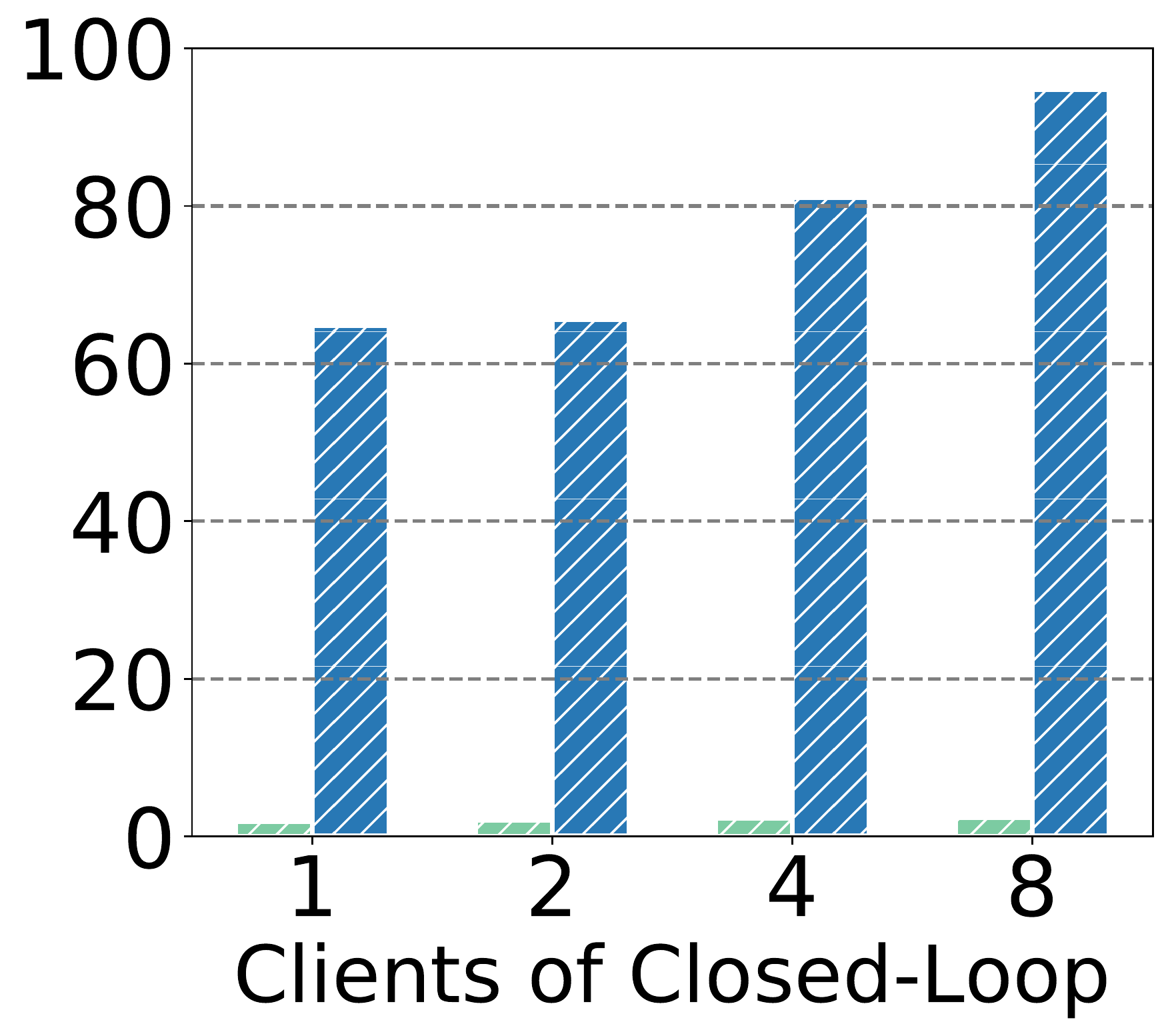}} \\ \vspace{-3mm}
  \hspace{0.5mm}
  \subfloat[$svd$]{\includegraphics[width=.225\textwidth]{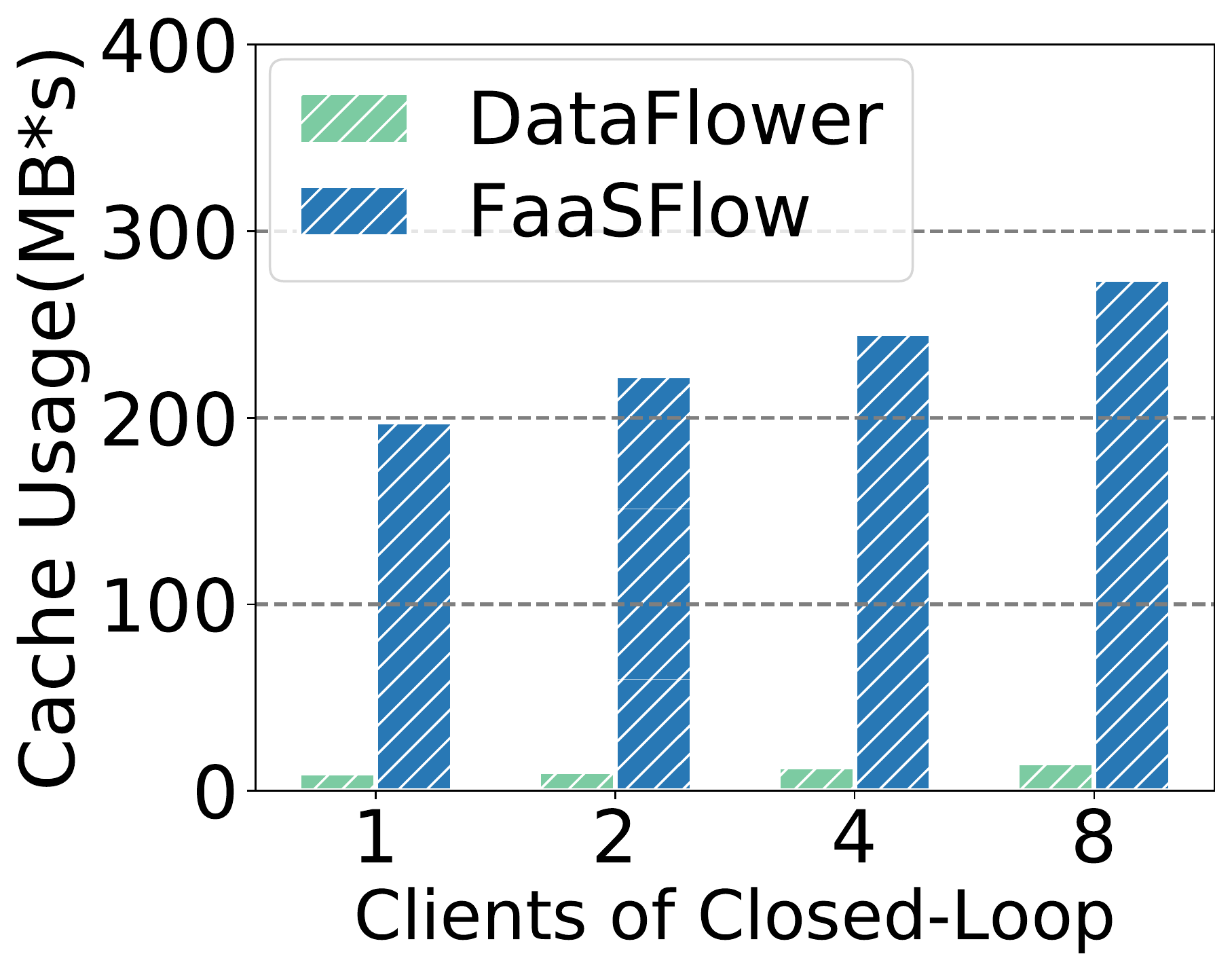}}
  \hspace{3.5mm}
  \subfloat[$wc$]{\includegraphics[width=.168\textwidth]{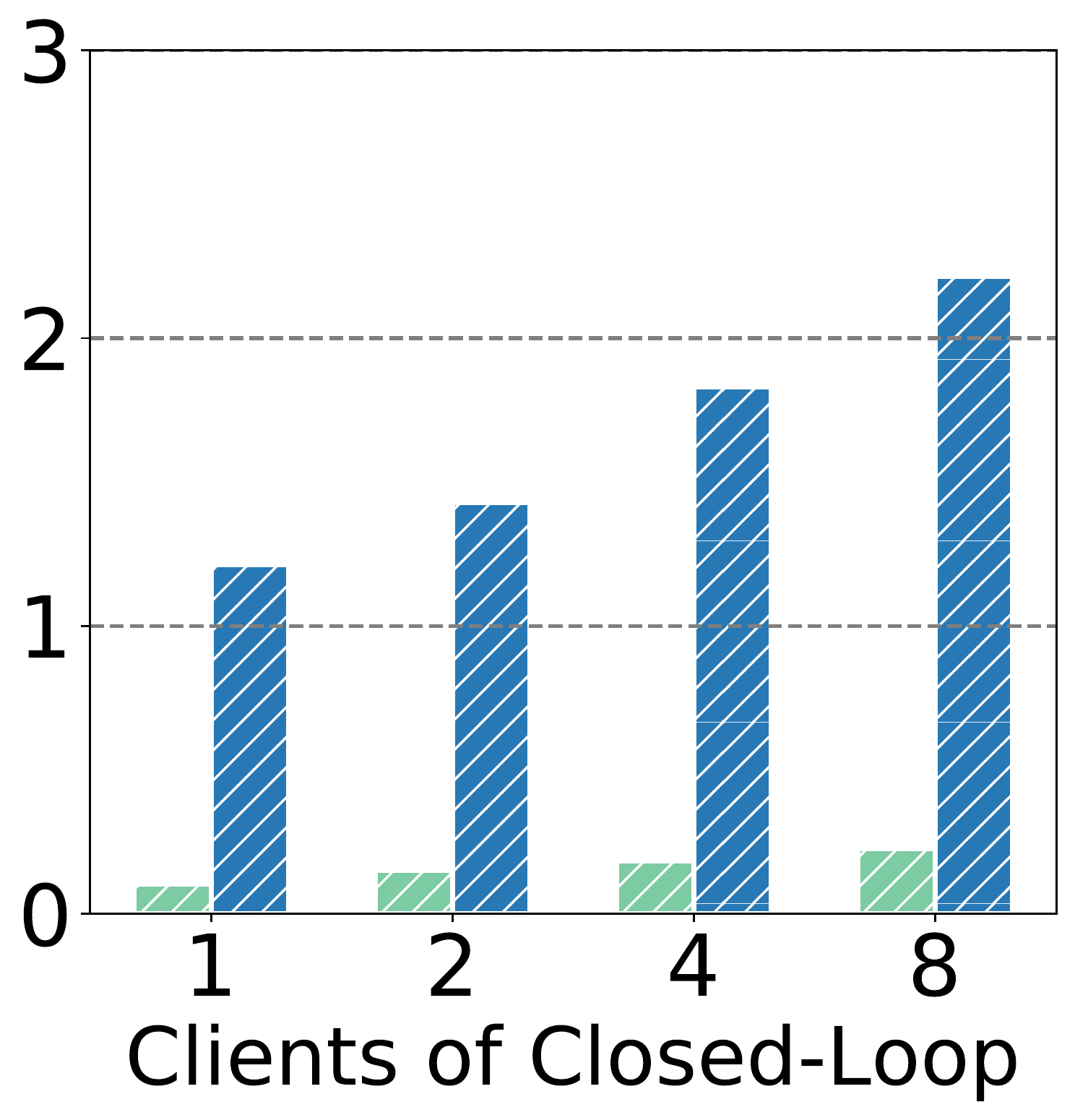}}\hspace{2mm}
  \caption{\label{fig:redis} The average memory usage on the host node for caching the intermediate data with DataFlower and FaaSFlow.}
\end{figure}

We also evaluate the effectiveness of the intermediate data management technique in the host-container collaborative communication mechanism of DataFlower. 
Figure~\ref{fig:redis} shows the average memory usage for caching the intermediate data on the host node per request with DataFlower and FaaSFlow at different loads. 
Compared to FaaSFlow, DataFlower reduces 19.1\%, 90.2\%, 94.9\% and 97.5\% of memory for caching the intermediate data in four benchmarks.
FaaSFlow uses large memory on the host node for the intermediate data, because it ignores the lifetime of the intermediate data with the control-flow paradigm. 
On the contrary, DataFlower recycles the memory space through proactive release and passive expire.

\subsection{Efficiency in Handling Bursty Load}
\begin{figure}
  \centerline{\includegraphics[width=.88\columnwidth]{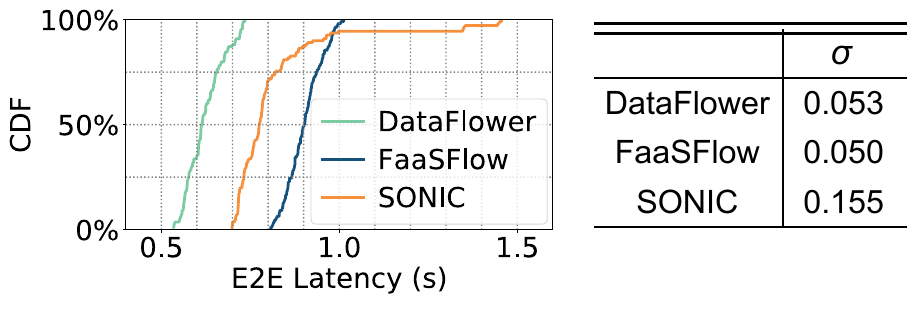}}
    \vspace{-1mm}
    \caption{\label{fig:loadburst} Response latency CDF and Standard Deviation ($\sigma$)}
\end{figure}
Given that serverless is more geared towards bursty workloads, this experiment evaluates the performance of DataFlower in the bursty case. 
In the experiment, we suddenly increase the load of the benchmark {\it wc} (from 10 rpm to 100 rpm) using asynchronous invocations.
Other benchmarks show similar result.
For the 110 user requests invoked in the two minutes, 
Figure~\ref{fig:loadburst} shows the cumulative distribution and the standard deviation ($\sigma$) of their latencies respectively.  


As observed, FaaSFlow and DataFlower handle bursty workloads more efficiently than SONIC. Moreover, the benchmarks have lower average and 99\%-ile latencies with DataFlower, compared with others.
DataFlower performs better, because it increases the resource utilization by overlapping the computation (FLU) and communication (DLU). 
In this way, each function container can execute more requests in a given time, and fewer containers are required to scale out for handling the bursty load.
Scale-out time is reduced and the overhead of creating containers with DataFlower is also reduced.




\subsection{Adaptiveness to Various Workflows}
\label{sec:stability}
\begin{figure}
  \captionsetup[subfloat]{margin=2pt,format=hang}
  \vspace{-3mm}
  \centering
  \subfloat[Fan-out and Fan-in branches]{\includegraphics[width=.21\textwidth]{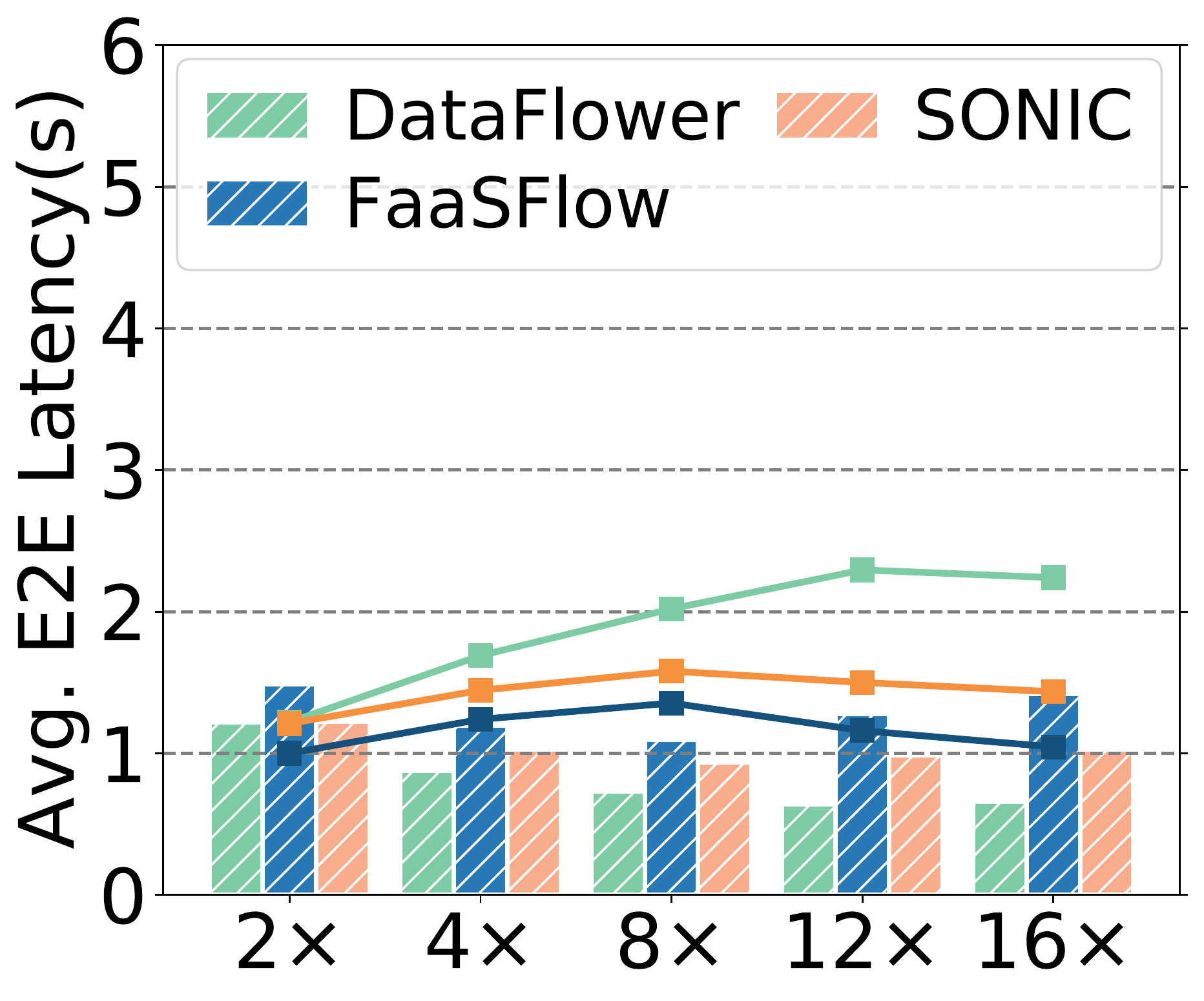}} \hspace{3mm}
  \subfloat[Input Datasize]
  {\includegraphics[width=.225\textwidth]{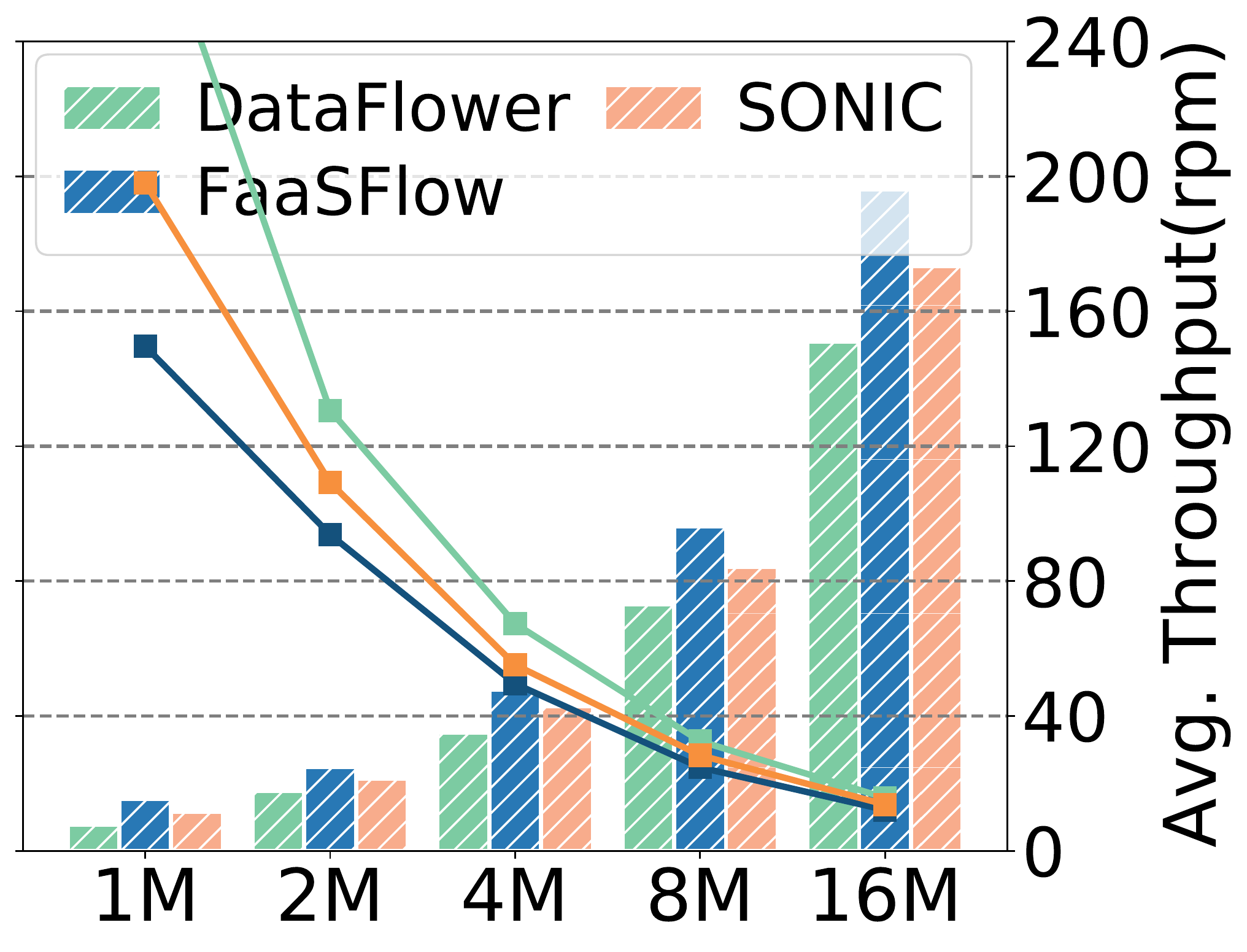}}
    \caption{\label{fig:sca} The average response latency and throughput of {\it wc} with different fan-out branches and input data size. 
  }
\end{figure}

While serverless workflows may have different structures and inputs, this experiment uses {\it wc} as an example to show the adaptiveness of DataFlower for various workflows. The DLU of the predecessor will send the data to child functions through different pipe connectors in the FIFO manner.
Figure~\ref{fig:sca} shows the average response latency and the processing throughput of {\it wc} with different fan-out and fan-in branches and input data sizes. The input data size is fixed to be 4MB in Figure~\ref{fig:sca}(a).
The number of fan-out branches is 4 in Figure~\ref{fig:sca}(b).

Observed from Figure~\ref{fig:sca}(a), DataFlower results in much lower latency and higher throughput of the benchmark with all the branch numbers. 
DataFlower increases the peak throughput by 69.3\% and 58.8\% compared with FaaSFlow and SONIC, respectively.
We can also find that DataFlower performs better with higher branch numbers.
When the fan-out branch increases, DataFlower can execute more functions in parallel. 
The data-availability driven per-function triggering in DataFlower takes advantage of the parallelism, the data is processed earlier and faster.
On the contrary, the sequential function triggering of the control-flow paradigm in FaaSFlow and SONIC fails to take advantage of the parallelism.

Observed from Figure~\ref{fig:sca}(b), the latency increases, and the throughput drops when the input data size increases with FaaSFlow, SONIC, and DataFlower. 
It is reasonable because a larger input data size means higher workload in a single request of {\it wc}.
With small input data (e.g. 1MB), DataFlower increases the throughput by 91.8\% and 44.9\% compared with FaaSFlow and SONIC.
With larger input data (e.g. 16MB), DataFlower increases the throughput by 29.5\% and 14.5\% compared with FaaSFlow and SONIC.
The improvement drops when the input data size increases.
With the given number of fan-out and fan-in branches, the workload of a function increases when the input data is larger. 
When the input data is large, the computation resources (i.e. CPU time) become the performance bottleneck, and the performance gain from the data-flow paradigm in DataFlower becomes smaller.

With the data-flow paradigm, it is beneficial to further reduce the function granularity. While the data-flow paradigm is capable of exploring parallelism, the fine-grained functions can run in parallel to explore the spare resources. On the contrary, the control-flow paradigm cannot support high parallelism and is therefore suitable for coarse-grained functions.


\subsection{Impact of Scaling Up Containers}
\label{sec:resource-scal}


When the number of functions that may run in parallel is small (e.g., the number of fan-out branches is small in {\it wc}), the response latency can not be reduced by increasing the containers (referred to be scaling out).
Increasing the resources allocated to each container may improve the performance in this case (referred to be scaling up).
\begin{figure}
  \centerline{\includegraphics[width=.9\columnwidth]{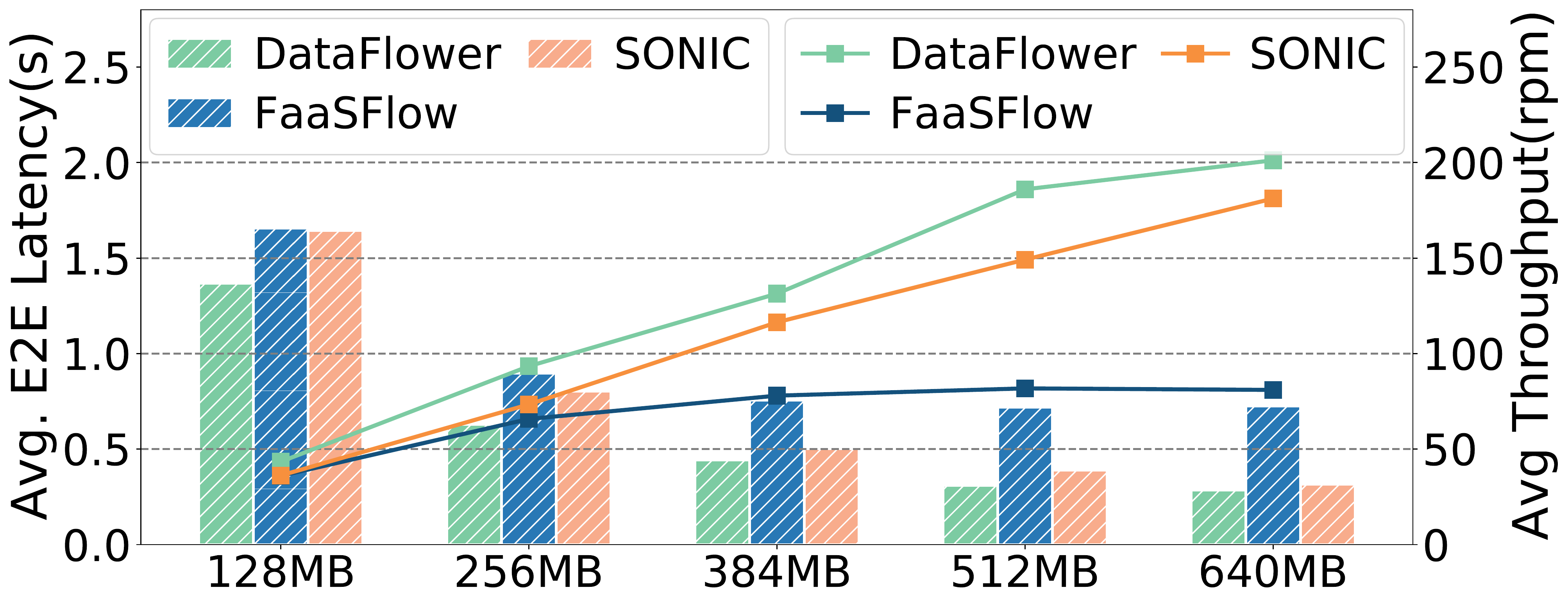}}
    \caption{\label{fig:sens} The response latency and throughput of $wc$ when scaling up the container's resources.}
\end{figure}

Figure~\ref{fig:sens} shows the response latency and the processing throughput of {\it wc} with the 4MB input and 8 fan-out branches, when we scale up the container's resource. The $x$-axis shows the memory specification allocated to a container. 
The CPU time and network bandwidth increase linearly with the memory allocation.
As observed, the processing throughput increases linearly when we scale up the containers with DataFlower and SONIC.
Compared with FaaSFlow and SONIC, DataFlower increases the throughput of {\it wc} by 148.4\% and 11.1\% with large containers (640MB memory).

When the container scales up, the performance of both function computation and the data transmission increase with DataFlower and SONIC, as they use direct data passing.
However, with FaaSFlow, the performance bottleneck occurs in accessing the backend storage when the functions store intermediate data in parallel.
In this case, FaaSFlow often cannot benefit from scaling up the containers.

Serverless platforms should support container scale-out and scale-up to support the higher load with the data-flow paradigm. The workflows with many fine-grained functions that may run in parallel tend to benefit more from the scale-out strategy, while the workflows with heavy functions also benefit from the scale-up strategy.



\subsection{Colocating Multiple Workflows}
\begin{figure}
  \captionsetup[subfloat]{margin=2pt,format=hang}
  \vspace{-3mm}
  \centering
  \subfloat[$img$]{\includegraphics[width=.214\textwidth]{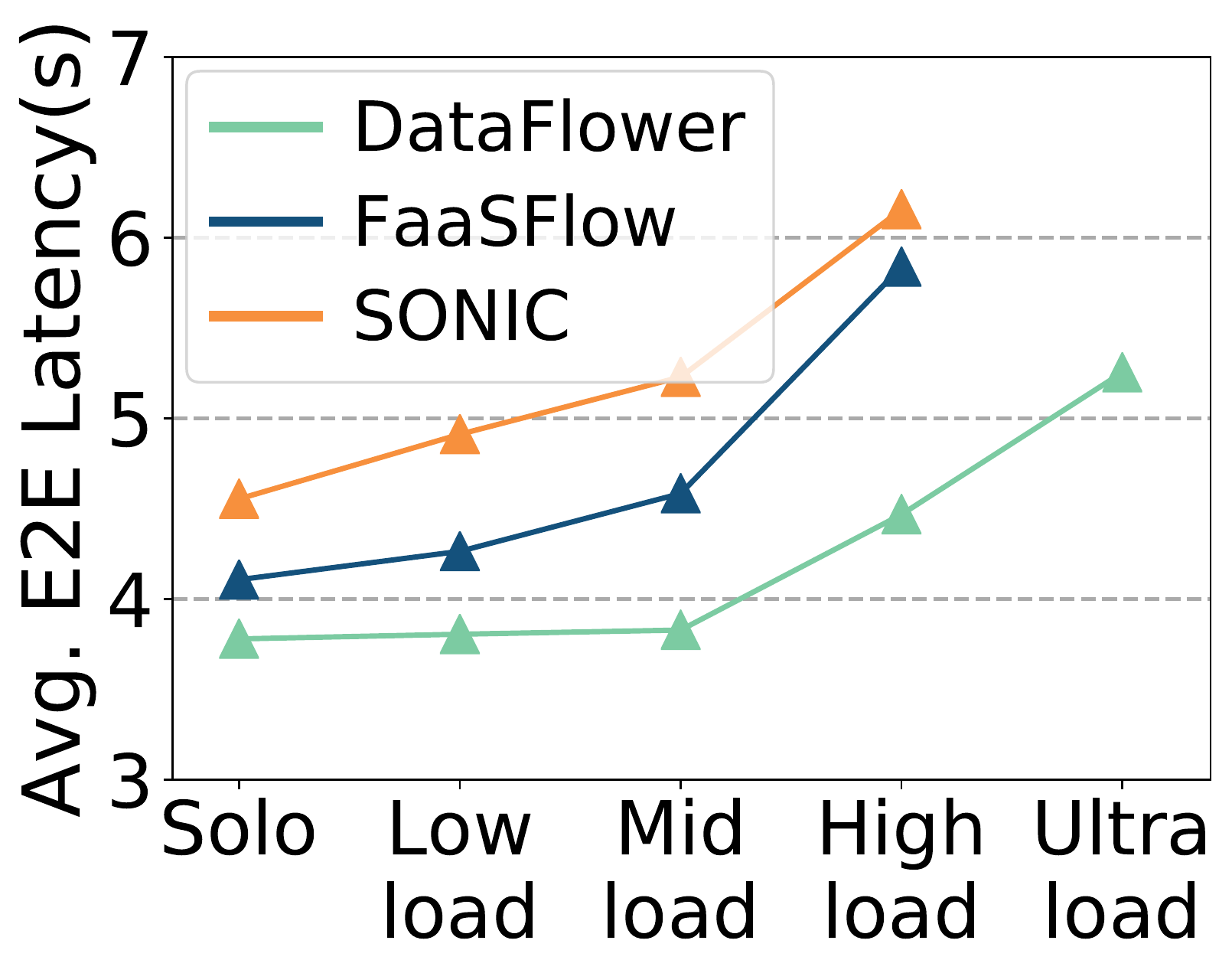}}
  \hspace{2mm}
  \subfloat[$vid$]{\includegraphics[width=.208\textwidth]{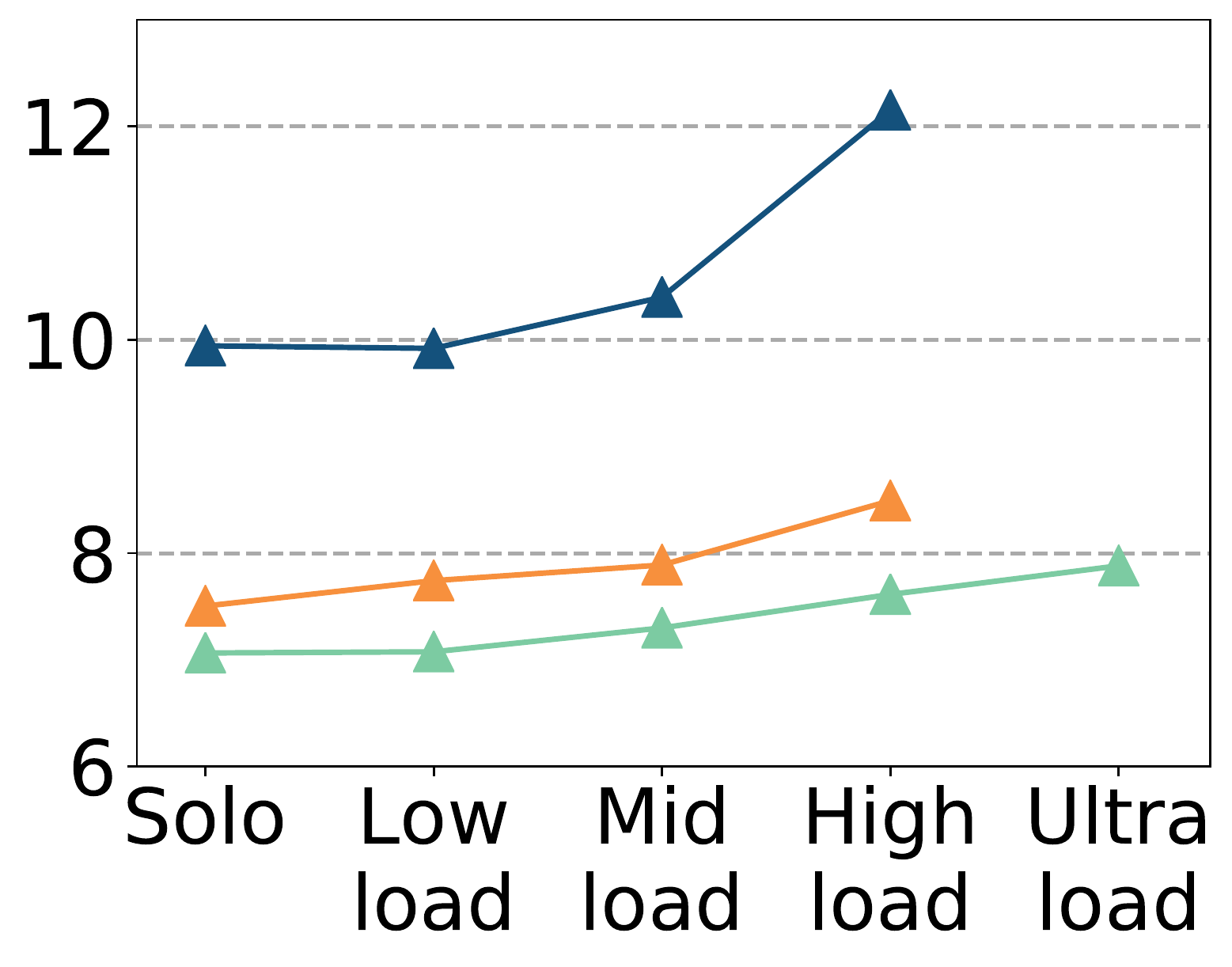}}
  \\ \vspace{-3mm}
  \subfloat[$svd$]{\includegraphics[width=.22\textwidth]{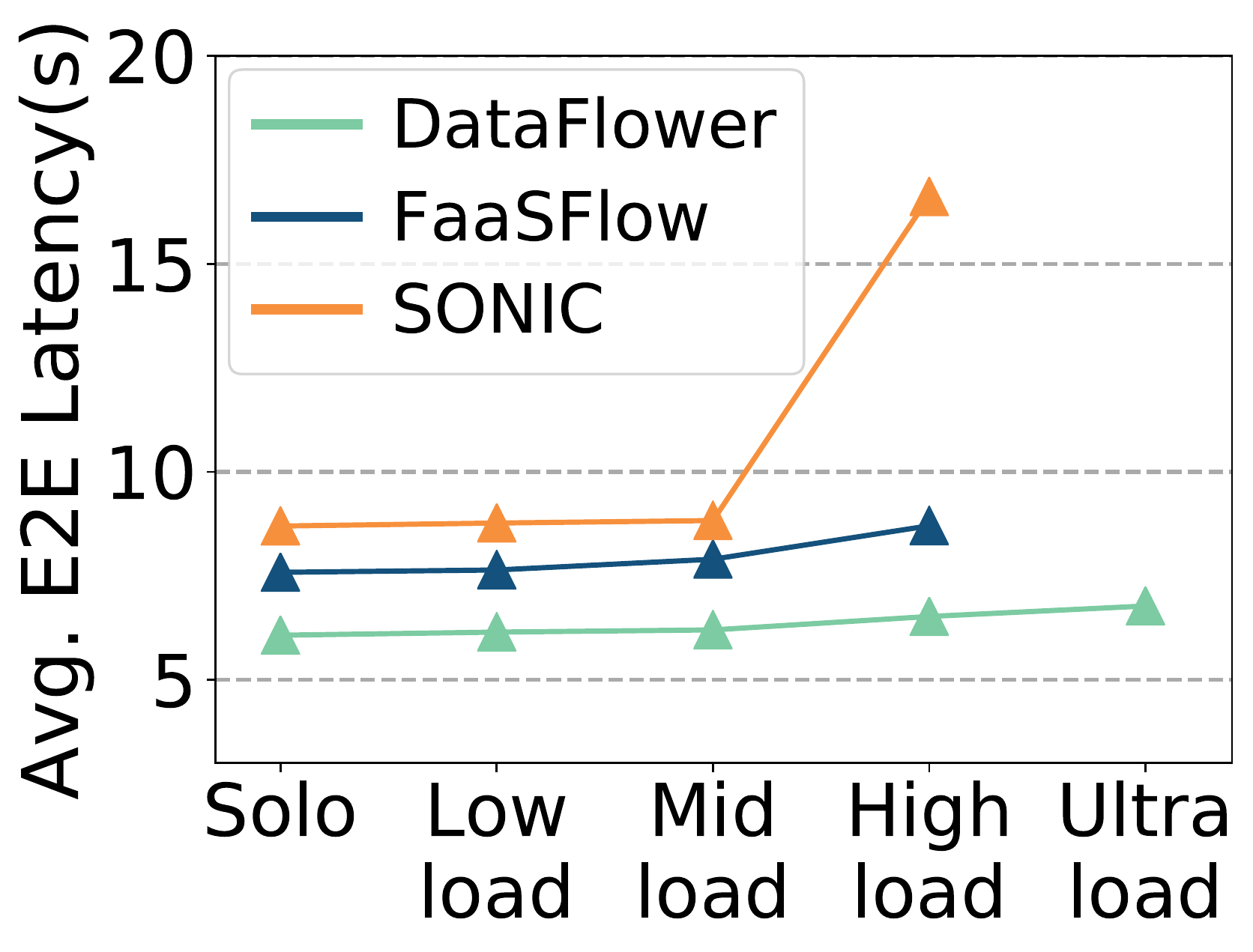}}
  \hspace{2mm}
  \subfloat[$wc$]{\includegraphics[width=.21\textwidth]{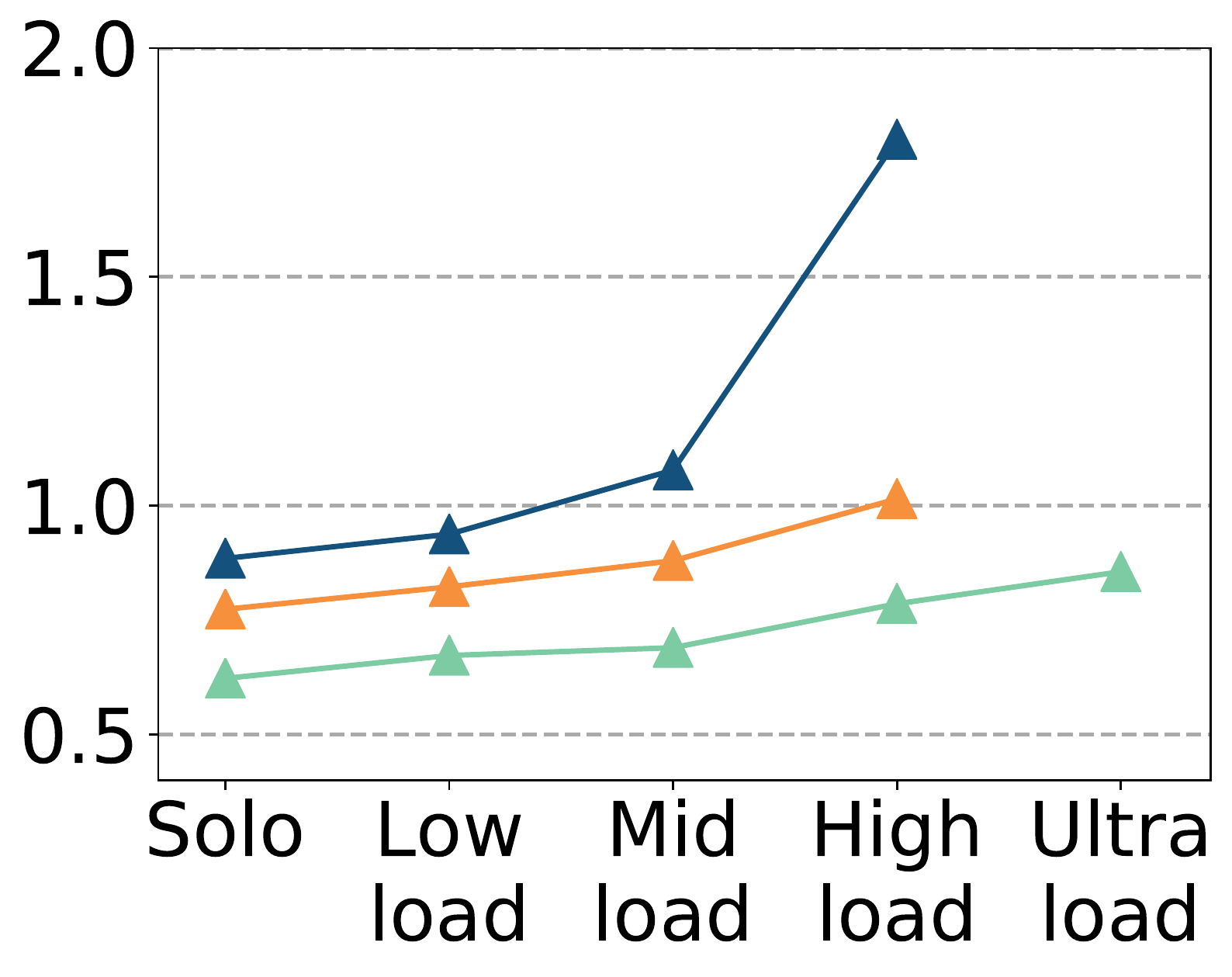}}
  \hspace{1mm}
  \caption{\label{fig:col} The response latency of the co-located benchmarks in DataFlower, FaaSFlow and SONIC.}
\end{figure}

In the production deployment, multiple serverless workflows may co-run on the same set of nodes.
In this experiment, we co-run all four benchmarks on the three worker nodes to evaluate DataFlower in the co-location scenario.
Figure~\ref{fig:col} shows the latencies of the co-located benchmarks. 
In the figure, ``Solo'' shows the benchmark's latencies when it runs alone,  
``Low load'' indicates when all four benchmarks run with low asynchronous workloads, 
``Ultra load'' shows the latencies when the loads of the benchmarks is maximized, in order to simulate the highly contended environments.

As observed, the benchmarks have the shortest response latency with DataFlower in all the co-location cases. The benchmarks fail at ``Ultra load'' with FaaSFlow and SONIC. 
This is because they lack the efficient container scaling policy on heavily overtaxed machines.
We can find that none of the benchmarks suffers from more than 2X performance degradation when the workload is high with DataFlower. 
The performance interference between workflows is limited, as the CPU resources are explicitly allocated to each container, and both FLU and DLU in a container only use its dedicated CPU, memory, and network resources.
Meanwhile, cloud providers are also working on encapsulating functions with MircoVM like Firecracker~\cite{DBLP:conf/nsdi/AgacheBILNPP20}, RunD~\cite{DBLP:conf/usenix/LiC0GBTZWHG22}, and Kata-container~\cite{kata}, to eliminate turbulent performance due to interference.



\subsection{Integrating with Stateful Functions}

\begin{figure}
  \centerline{\includegraphics[width=.88\columnwidth]{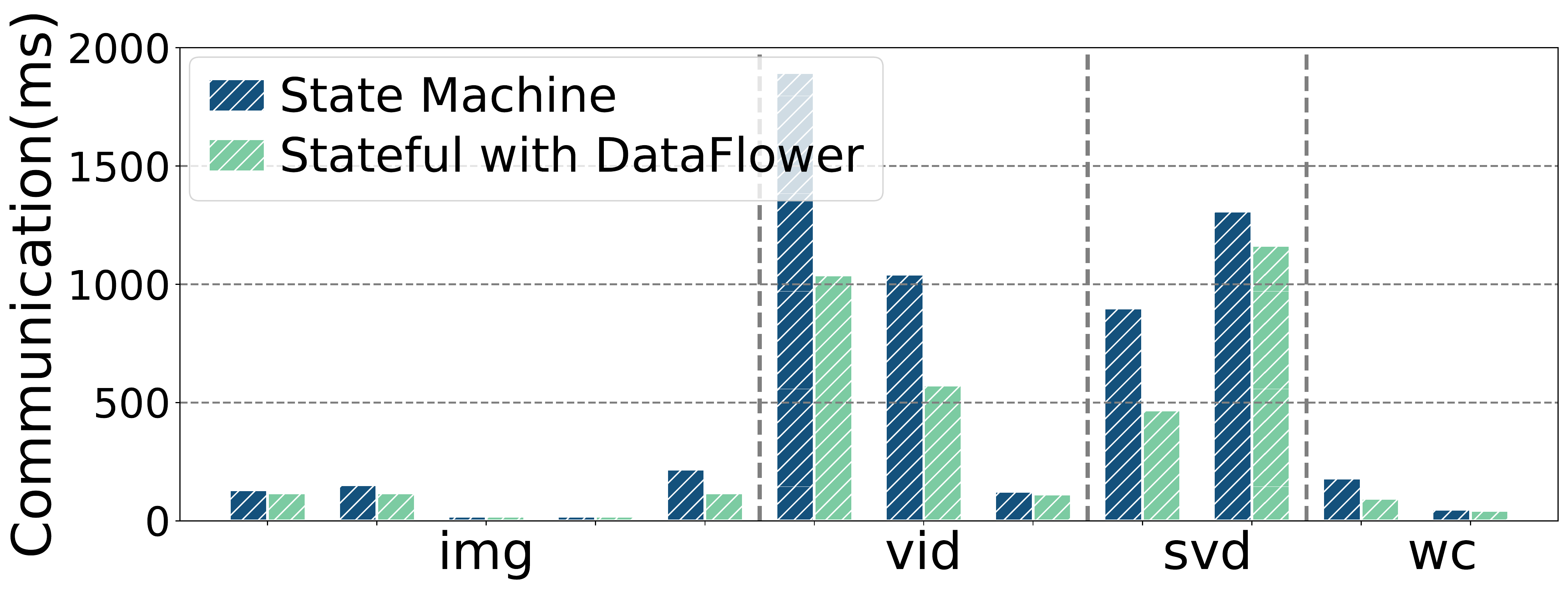}}
     \caption{\label{fig:streaming_vs_aws} The communication overhead with the traditional state machine and streaming-based functions on DataFlower.}
\end{figure}

DataFlower can also be extended and applied to workflows using stateful functions. For example, AWS Step Functions~\cite{StepFunctions} implements stateful serverless workflows by using ``state machine'' to schedule tasks~\cite{awsstate}. 
Specifically, when a Lambda function $\lambda_a$ completes, it returns result and the state machine stores the output as a context object. Then the state machine advances to the next state and calls the next Lambda function $\lambda_b$ by passing the data from the context object.
Since the built-in state machine in AWS Step Functions has a quota of 256KB for stateful data~\cite{stepquota}, we simulate the situation where benchmarks have been deployed in a stateful manner with a state machine running on AWS EC2 to cache unlimited data.

Figure~\ref{fig:streaming_vs_aws} shows the data transfer time of the functions in each benchmark with the above stateful function implementation. 
``Stateful with DataFlower'' shows the case that we use the streaming-based functions in DataFlower to replace the traditional state machine-based Step Functions.
As observed, the pipe connector in DataFlower reduces up to 47.6\% of the function-to-function data transfer time.

Moreover, the compute-communication overlap and data availability-based early triggering techniques are not affected by whether the functions are stateless or stateful. The experiments in Section~\ref{sec:early} have already showed their effectiveness. 
Therefore, DataFlower scheme can also improve the performance of serverless workflows based on stateful functions.



\section{Conclusions and Future Work}
We propose DataFlower, a scheme that achieves the data-flow paradigm for serverless workflows. 
Specifically, the container that runs a function is abstracted into a FLU and a DLU. The FLU processes the function computation, and the DLU handles all the communication. 
In this way, the computation and communication overlap with each other, improving the processing throughput.
Moreover, a host-container collaborative communication mechanism is proposed 
to minimize the communication overhead. 
Experimental results show that DataFlower greatly reduces the response latency, increases the supported throughput, and reduces resource usage.


The data-flow paradigm provides an alternative way to prewarm containers based on the data dependencies and availability. With the prior knowledge of the data dependencies, we are designing a policy to warm up a container for a function based on the data-availability instead of predicting function execution patterns in the future work.


\bibliographystyle{ACM-Reference-Format}
\bibliography{references}










\end{document}